\documentclass[pra,twocolumn, floatfix, footinbib,superscriptaddress, longbibliography]{revtex4-2}
\usepackage[utf8x]{inputenc}
\usepackage[english]{babel} 
\usepackage[english]{babel} 
\usepackage[colorlinks=true,bookmarks=false,citecolor=blue,urlcolor=blue]{hyperref}

\usepackage{amsmath,amssymb,esint}
\usepackage{graphicx}
\usepackage{graphics}
\usepackage{verbatim}

\usepackage{bbold}
\usepackage[english]{babel}
\usepackage{bm}
\usepackage{natbib}
\usepackage[normalem]{ulem}
\usepackage{epstopdf}
\usepackage{multirow}
\usepackage{ dsfont }
\usepackage{tikz}
\DeclareMathAlphabet\mathbfcal{OMS}{cmsy}{b}{n}

 \usepackage{booktabs}
 \usepackage{multirow}
\usepackage{ragged2e}

\usepackage{amsmath}
\usepackage[colorinlistoftodos]{todonotes}
\usepackage{color}

\newcommand{\bee}{\begin{equation}}
\newcommand{\ene}{\end{equation}}

\newcommand{\bea}{\begin{eqnarray}}
\newcommand{\ena}{\end{eqnarray}}

\definecolor{persianblue}{rgb}{0.11, 0.22, 0.73}

\newcommand{\ve}[1]{{\bf {#1}}}

\graphicspath{{pics/}}

\let\vec=\mathbf

\renewcommand {\phi}{\varphi}

\newcommand{\vphi}{\varphi}
\let\vec=\mathbf

\usepackage[normalem]{ulem}

\begin{document}
\title{Nonlinear circular dichroism in Mie-resonant nanoparticle dimers }
\author{Kristina Frizyuk}
\email{k.frizyuk@metalab.ifmo.ru}
\affiliation{Department of Physics and Engineering, ITMO University, St Petersburg 197101, Russia}

\author{Elizaveta Melik-Gaykazyan}
\affiliation{Nonlinear Physics Centre, Research School of Physics, Australian National University, Canberra ACT 2601, Australia}
\author{Jae-Hyuck~Choi}
\affiliation{Department of Physics, Korea University, Seoul 02841, Republic of Korea}
\author{Mihail Petrov}
\affiliation{Department of Physics and Engineering, ITMO University, St Petersburg 197101, Russia}
\author{Hong-Gyu~Park}
\affiliation{Department of Physics, Korea University, Seoul 02841, Republic of Korea}
\affiliation{KU-KIST Graduate School of Converging Science and Technology, Korea University, Seoul 02841, Republic of Korea}
\author{Yuri Kivshar}
\email{yuri.kivshar@anu.edu.au}
\affiliation{Nonlinear Physics Centre, Research School of Physics, Australian National University, Canberra ACT 2601, Australia}


\begin{abstract}
We study nonlinear response of a dimer composed of two identical Mie-resonant dielectric nanoparticles illuminated normally by a circularly polarized light. We develop a general theory describing {\it hybridization of multipolar modes} of the coupled nanoparticles, and reveal nonvanishing {\it nonlinear circular dichroism} (CD) in the second-harmonic generation (SHG) signal enhanced by the multipolar resonances in the dimer provided its axis is oriented under an angle to the crystalline lattice of the dielectric material. We present experimental results for this SHG-CD effect obtained for the AlGaAs dimers placed on an engineered substrate which confirm the basic prediction of our general multipolar hybridization theory. 
\end{abstract}
\maketitle



Many phenomena in nature, including multiple biochemical processes, are governed by the fundamental property of chirality.  An object is called chiral when its mirror image cannot be superimposed with the original object, and many examples of chirality can be found at all scales in nature, from organisms to biomolecules and amino acids, which often occur only in one handedness. {\it Circular dichroism spectroscopy} was suggested as a powerful optical technique for the study of chiral materials and molecules. It gives access to an enantioselective signal based on the differential absorption of right and left circularly polarized light \cite{Nakanishi1995,Greenfield_2006,Hopkins2016-Circulardichro,DeSilva2021-UsingCircularDichro}. 

In natural media, chiral effects are generally weak, and chiral plasmonic structures~\cite{PhysRevX.2.031010,Na_Liu2017,Valev2017,Li2021-TunableChiralOptics} and chiral metamaterials~\cite{Decker_Klein_Wegener_Linden_2007,Wang_Cheng_Winsor_Liu_2016,Zhao2012-Twistedopticalmetam} were suggested as a new tool for achieving strong chiroptical responses. Chiroptical activity is commonly quantified as a linear effect being characterized in terms of the so-called {\it circular dichroism} (CD). Most of the studies of CD effects in metamaterials have been focused on the scattering properties including reflection and transmission, and such effects attracted considerable attention due to their potential applications for biological sensing~\cite{dionne}, spin-sensitive resonant transmission~\cite{maxim}, and chiral meta-holograms~\cite{tao_li}. Also, chiroptical effects have been studied in systems of a few reconfigurable chiral meta-molecules {using metallic and dielectric particles as artificial building blocks~\cite{Krasnok2019,Banzer2019,Nechayev2018-ChiralityofSymmetr,Bautista2012-Second-HarmonicGener}.}
 
Circularly polarised light interacting with resonances of nanostructures in the nonlinear regime and nonlinear chiral metasurfaces can become extremely sensitive to slight asymmetries and thus give rise to a circular dichroic signal that is order of magnitude higher with respect to the same signal obtained with the fundamental frequency~\cite{PhysRevLett.107.257401,Hooper_Mark_Kuppe_Collins_Fischer_Valev_2017,Belardini2020-CircularDichroismin,Rodrigues2014-NonlinearImagingand,Bertolotti2015-Secondharmoniccircu,Schmeltz2020-Circulardichroismse}. One of the first studies of {\it nonlinear circular dichroism} in the second-harmonic generation (SHG) signal was reported for chiral nanostructures consisting of $G$-shaped elements made of gold~\cite{Valev_2009,PhysRevLett.104.127401}, and most recently for the third-harmonic generation in a planar system of shifted gold bars~\cite{hayk}. It was shown that in such plasmonic structures the nonlinear signal originates from “hot spots” being dependant on the handedness of the nanostructures~\cite{Tymchenko2015-GradientNonlinearPa}. 

Although metal nanostructures were explored first for nonlinear chiral nanophotonics, high-index dielectric nanoparticles can sustain strong magnetic resonances and higher-order Mie-type modes in the visible and near-infrared frequency range~\cite{Koshelev_2020}, and they are highly suitable for nonlinear nanophotonics \cite{Smirnova_Kivshar_2016,Frizyuk_Volkovskaya_Smirnova_Poddubny_Petrov_2019,Gigli2019-All-DielectricNanore,Renaut2019-ReshapingtheSecond} and nonlinear chiral metasurfaces~\cite{Werner2020} underpinning a strong nonlinear multipolar response, in contrast to the metal structures where the resonant fields are tightly confined to the surface.

Semiconductor compounds and alloys of the III-V group such as AlGaAs often exhibit a large second-order susceptibility because of a lack of inversion symmetry of their crystal lattice. Consequently, efficient SHG has been observed in isolated semiconductor nanoantennas, being enhanced by Mie resonances, optical anapoles, or bound states in the continuum \cite{Sanatinia_Anand_Swillo_2014,Carletti2016-ShapingtheRadiation,Kruk_Camacho-Morales_Xu_Rahmani_Smirnova_Wang_Tan_Jagadish_Neshev_Kivshar_2017,Timofeeva_Lang_Timpu_Renaut_Bouravleuv_Shtrom_Cirlin_Grange_2018,Sautter2019-TailoringSecond-Harm,Koshelev_Kruk_Melik-Gaykazyan_Choi_Bogdanov_Park_Kivshar_2020}.

\begin{figure}[ht!]
  \includegraphics[width=0.75\linewidth]{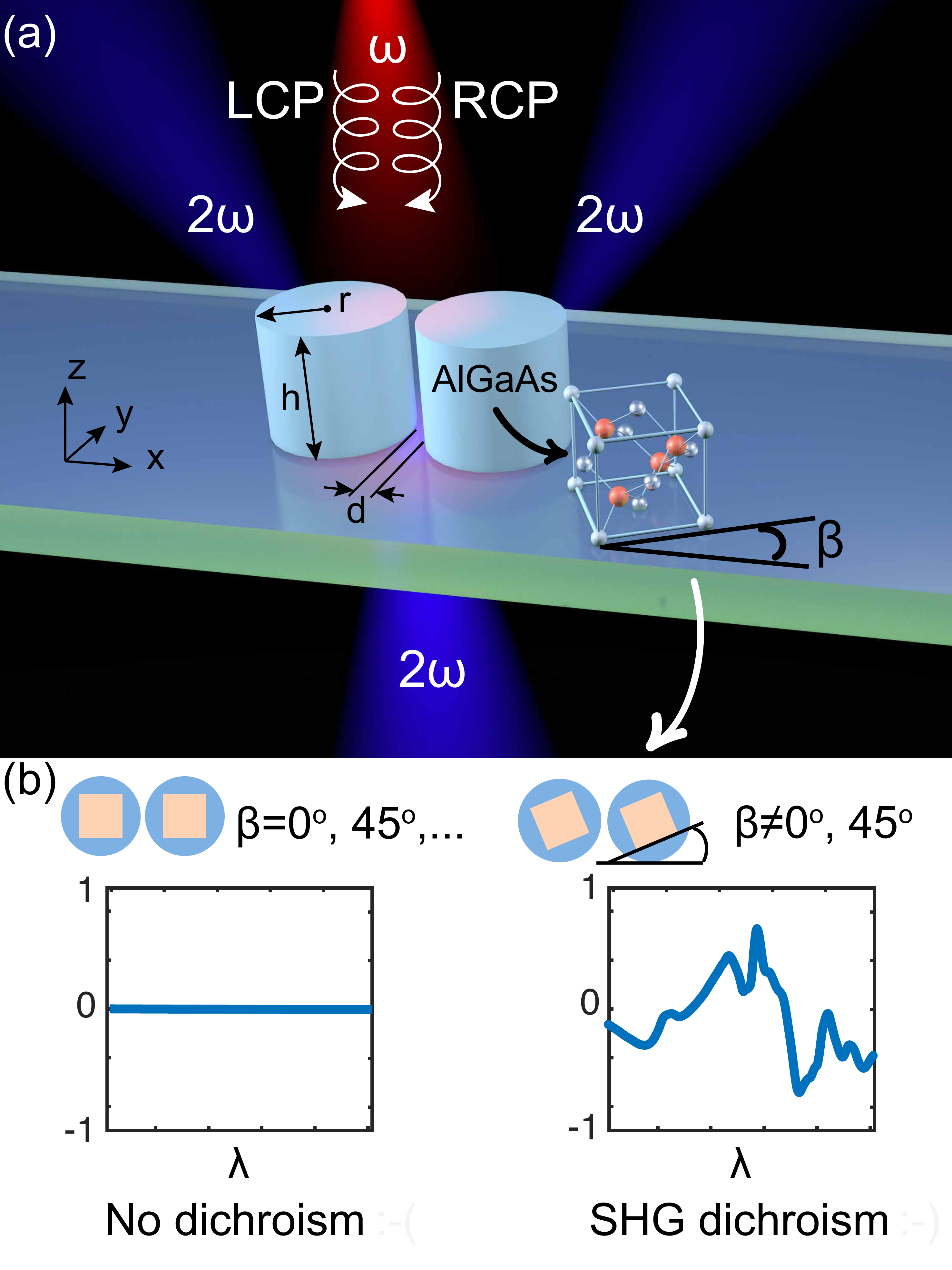}
\caption{(a) Concept of nonlinear circular dichroism in Mie-resonant nanoparticle dimers. Dimer axis is parallel to the $x$-axis, $\beta$ - angle between AlGaAs crystalline axis [100] and the $x$-axis. (b)  Asymmetric SHG driven by a circularly polarized light is generated from a dimer with an arbitrarily oriented crystalline lattice. For other orientations such as $[100]||x$ and $[110]||x$, nonlinear circular dichroism is not observed.}
\label{fig:concept}
\end{figure}

    In this Letter, we develop the general hybridization theory for multipolar Mie-resonant modes and then study nonlinear response of a nanoparticle dimer (shown in Fig.~\ref{fig:concept} (a)) illuminated normally by a circularly polarized light. We  observe, for the first time to our knowledge, that nonvanishing circular dichroism of the second-harmonic (SH) signal may occur due to the interplay of multipolar magnetic and electric Mie resonances provided the dimer axes is oriented under an angle to the crystalline lattice of the dielectric material (Fig.~\ref{fig:concept} (b)). We present a general formalism and modal theory to explain the origin of this unexpected phenomenon, and also verify our general multipolar hybridization theory in the experiment for the Mie-resonant SHG from the AlGaAs dimers placed on an engineered substrate.

\section{Results and Discussion}

We consider a dimer structure, which consists of two AlGaAs cylinders with height $h=635$ nm and radius $r=475$ nm, placed on an engineered substrate, 300-nm ITO on glass
with an added SiO$_2$ spacer  350-nm thick, similar to the structures studied earlier for the observation of quasi-bound states in the continuum for isolated nanoantennas~\cite{Koshelev_Kruk_Melik-Gaykazyan_Choi_Bogdanov_Park_Kivshar_2020}. 
Distance between two cylinders $d=30$ nm. Incident plane wave is circularly polarized  propagating in the negative direction of  the z-axis (see  Fig.~\ref{figlin} (b)), while the SH power is collected in the upper media. We show the difference between left and right circular polarized incident waves (LCP and RCP, respectively) appears  in the pattern of  second-harmonic radiation generated by a single  particle, while overall intensity of SHG stays the same. At the same time, for a dimer structure one can achieve difference in the integral SHG intensity as well, but only under proper orientation between the  crystalline lattice of the dielectric material and dimer axis. 

First, we develop a general theory that demonstrates the origin of nonlinear CD in nanoparticle dimers. We write the SH polarization as follows \cite{Boyd2003}:
\begin{equation}
 \ve P^{2\omega} (\vec r)=\varepsilon_0 \hat\chi^{(2)} \vec E^{inc} (\vec r)\vec E^{inc} (\vec r)\label{pol},
\end{equation}
where $\vec E^{inc}(\vec r)$ is the fundamental field inside the nanoparticle. 
\begin{widetext}
To obtain SH field, we employ the dyadic Green function formalism \cite{Frizyuk_Volkovskaya_Smirnova_Poddubny_Petrov_2019,Novotny}
\begin{equation}
\mathbf{E}^{2\omega}({\mathbf{r}})=(2\omega)^2\mu\int\limits_V dV' \hat {{\bf G}}({\bf r,r'},k) \mathbf{P}^{2\omega} (\mathbf{r}')=(2\omega)^2\mu\int\limits_V dV' \sum_{n} \frac{\mathbf{E}_{n}(\mathbf{r}) \otimes \mathbf{E}_{n}\left(\mathbf{r}^{\prime}\right)}{2 k\left(k-k_{n}\right)} \mathbf{P}^{2\omega} (\mathbf{r}')\label{shfield},
\end{equation}
\begin{figure}[ht!]
 \includegraphics[width=0.75\linewidth]{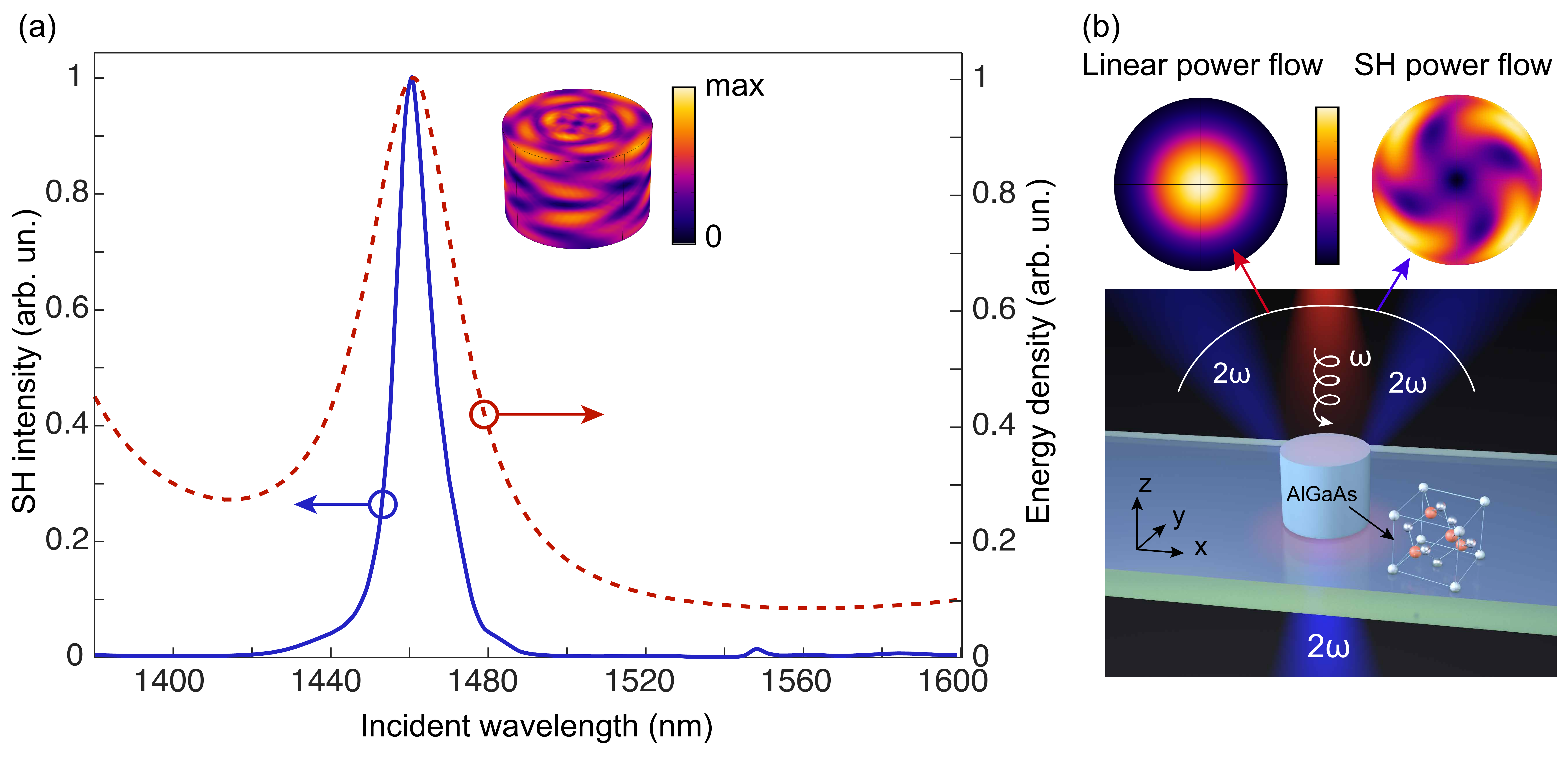}
\caption{SHG from an isolated dielectric cylinder. (a) Energy density (red dashed curve) and SH intensity into upper half-space (blue solid curve). Field amplitude at the cylinder surface is shown at the resonance. (b) Linear {scattered} and nonlinear power flows into upper {half-space} shown for $\lambda=1460$nm. Linear power flow has rotational symmetry, while nonlinear  has $C_4$ symmetry. The color-bar is presented in linear color-scale. }\label{figlin}
\end{figure}
\end{widetext}
where we employ the resonant-state expansion of the Green function \cite{PhysRevA.90.013834},  and calculate the integral over the particle's volume, $\mathbf{E}_{n}\left(\mathbf{r}\right)$ being system's eigenstate. The key role in our results is played by the overlap integral\cite{Gigli2020-Quasinormal-ModeNon} that appears in Eq.~\eqref{shfield} as follows, 
\begin{equation}
D_n=\int\limits_V dV' \mathbf{E}_{n} \left(\mathbf{r}^{\prime}\right) \mathbf{P}^{2\omega} (\mathbf{r}')\label{overlap}
\end{equation}

To calculate the overlap integral, we need to know both polarization and mode content in two cases--a single nanoparticle and a dimer, and below we consider these two cases subsequently.

\subsection{SHG for a single nanoparticle} 

{We performed numerical modeling with COMSOL Multiphysics\texttrademark \  simulation software.} The spectral dependence of  the {SH power flow emitted by a single dielectric resonator integrated over upper half space} is shown in Fig.~\ref{figlin}(a) by the blue solid curve. The spectrum of electromagnetic energy of the fundamental field inside the nanoparticle is shown by a red dashed curve. The SH intensity is maximal when the particle has a resonance at both fundamental and SH wavelengths. The inset of  Fig.~\ref{figlin}(a) shows the distribution of SH field at the nanoparticle surface at the resonant wavelength corresponding to excitation of a Mie mode, which properties will be discussed later. The   energy flux of the linear scattered field in the upper half-space is depicted in   Fig.~\ref{figlin}(b) along with the SH power flux. One can notice that the linear scattering pattern has a cylindrical symmetry whereas the SH radiation pattern has $C_4$ symmetry characterized by four petals. The rotation angle and shape of the petals depend on the wavelength. If we change the incident polarization from LCP to RCP, this four-lobe radiation pattern will be reflected in a mirror plane coinciding with the (100) crystalline plane cut through the center of the nanoparticle. 

The particular shape of the radiation patterns can be explained in terms of the multipolar decomposition of the eigenmodes~\cite{Gladyshev_Frizyuk_Bogdanov_2020,Brandl_Mirin_Nordlander_2006}. Multipoles $\vec W_{pm\ell}$ include both {\it electric multipoles} $\vec N_{pm\ell}$ and {\it magnetic multipoles} $\vec M_{pm\ell}$, which are defined  in the previous work by~\cite{Gladyshev_Frizyuk_Bogdanov_2020}. The symmetry group of a single cylindrical nanoparticle is $C_{\infty v}$, and its eigenmodes correspond to irreducible representations of this group~\cite{Zheng2015-OntheUseofGroupT,Chikkaraddy2017-HowUltranarrowGapS,Gladyshev_Frizyuk_Bogdanov_2020,Xiong_Xiong_Yang_Yang_Chen_Wang_Xu_Xu_Xu_Liu_2020}, while mode refers to a particular azimuthal number $m$, see Fig.~\ref{fig3}(b).   

In order to provide a simpler physical picture, we will operate in terms of scalar spherical functions here rather than  vector functions as they have the same symmetry, which is important for explaining the observed effects. However, the more strict analysis in terms of vector functions is provided in Supporting Information file. We use the cylindrical coordinate system corresponding to the $z$-axis, and  the incident circularly polarized wave has  $e^{\pm i \varphi}$ dependence. Here and below, the upper sign "$+$"refers to the LCP fields and lower "$-$" to the RCP fields. Under the normal illumination, 
only modes with $|m|=1$ are excited ({marked with the red tick in Fig.~\ref{fig3}(b)}). Thus, the fundamental field inside the nanoparticle can be written in the form,  $E^{inc}(r,z,\varphi)\propto E_{0}^{inc}(r,z) e^{\pm i \varphi}$. 

Rewriting the tensor $\hat\chi^{(2)}$ in the cylindrical coordinates, one can make sure that   it  behaves as $\sin(2\varphi)$. 
Thus, after substituting the results into Eq.~\eqref{pol}, we can write 
$ P^{2\omega} (r, z, \varphi)\propto  \sin(2\varphi)\exp(\pm 2 i \varphi) \propto 1-\exp(\pm 4i\varphi).$ 
\begin{widetext}
If we rotate the crystalline lattice by the angle $\beta$ as shown in Fig.~\ref{fig:concept} (b), the polarization is transformed to the form
 \begin{equation}P^{2\omega} (r, z, \varphi)\propto \sin[2(\varphi-\beta)] \exp(\pm 2 i \varphi)\propto 1-\exp[\pm 4i(\varphi-\beta)].\end{equation}
 Similar result is obtained in Supporting Information in the framework of more rigorous vectorial analysis. A nonvanishing overlap integral Eq.~\eqref{overlap} involves only the modes for which $m=0$ and $m=\pm 4$ ({Fig.~\ref{fig3}(b), blue ticks}). 

With help of Eq.~\eqref{shfield}, the SH field then can be decomposed into two terms with according  angular dependence:
\begin{equation}
{E}^{2\omega}=\sum_{n}  g_n(\omega)  E_{0n}^{2\omega}(r,z)+\sum_{\nu}q_\nu(\omega) E_{4\nu}^{2\omega}(r,z)e^{\pm 4i(\varphi-\beta)}\propto a(r,z,\omega)+b(r,z,\omega) e^{i\alpha(r,z,\omega) \pm 4i(\varphi-\beta)}
\label{eq:SHdecomp}
\end{equation}
where $n, \nu$ are the mode indices, $E_{0n}^{2\omega}$ and $E_{4\nu}^{2\omega}$  are the fields of the corresponding eigenmodes being proportional to the overlap integral. While  $g_n(\omega)$ and $q_{\nu}(\omega)$ are complex coefficient, the coefficients $a(r,z,\omega)$ and $b(r,z,\omega)$ are real-valued, and the function $\alpha(r,z,\omega)$ describes the relative phase.

\begin{figure}[ht!]
  \includegraphics[width=0.85\linewidth]{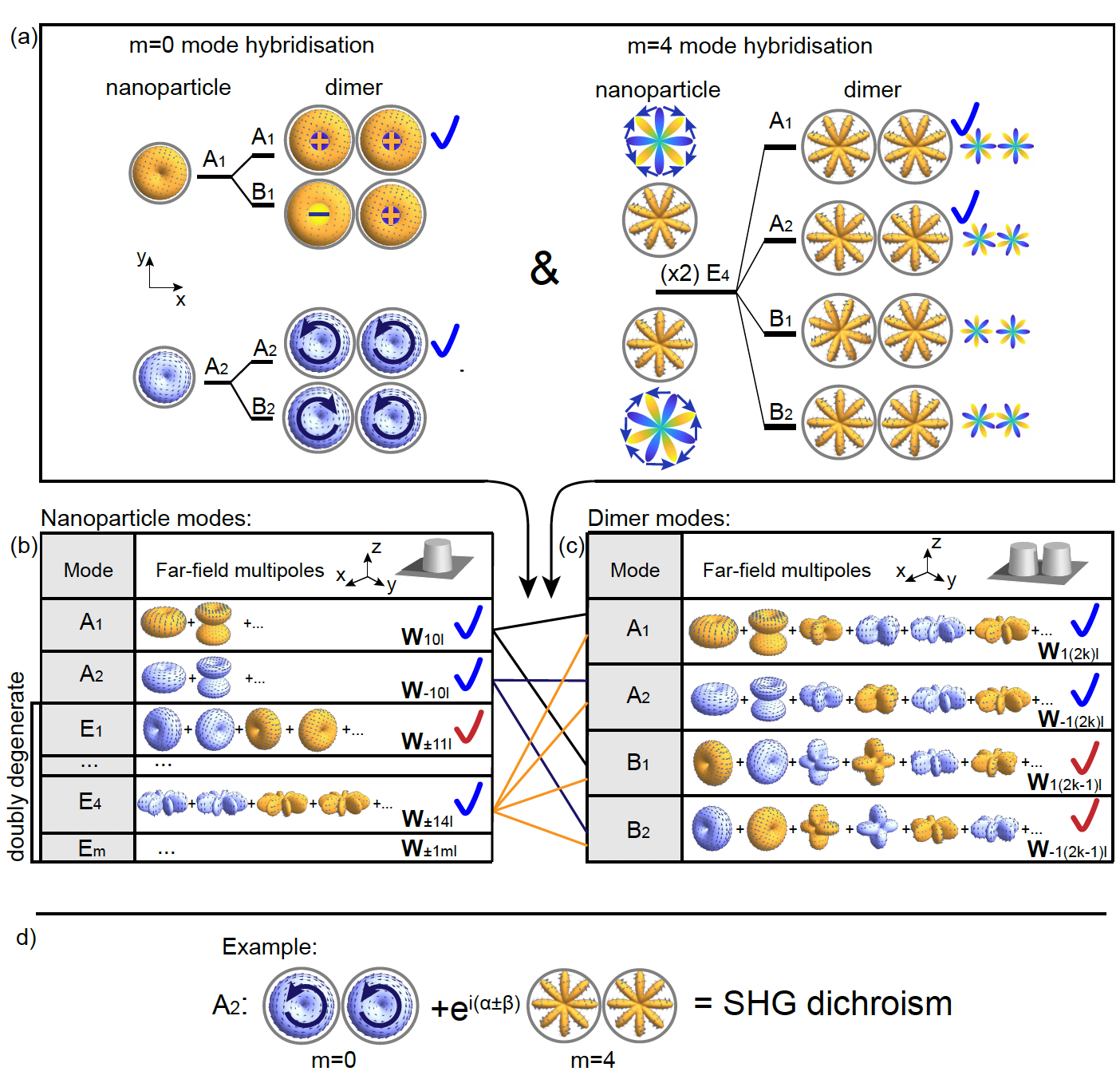}
\caption{ Hybridisation of the multipolar dimer modes. (a) Hybridisation is shown for the modes with $m=0$ (non-degenerate) and $m=4$ (doubly degenerate) in terms of field inside the particles. Scalar yellow-blue functions have the same symmetry as vector spherical harmonics, and they are shown as a guide for an eye. Modes excited in SHG are shown by the blue ticks, and in linear scattering - by the red ticks. (b) Multipolar content of the single cylindrical nanoparticle's modes. (c) Multipolar content of the dimer modes far-field. (d) Both m=0 and m=4 contribute into the same radiation channel, resulting in SH dichroism.}
\label{fig3}
\end{figure}
\end{widetext}
The intensity depends on $\varphi$ as 

$|E^{2\omega}|^2\propto |a+b e^{i\alpha}e^{\pm4i(\varphi-\beta)}|^2$
$=a^2+b^2+2ab\cos[\pm4(\varphi-\beta)+\alpha)]$, which results in appearance of four petals in  Fig.~\ref{figlin}(b). {Note, that both $m=0$ and $m=4$ modes play role in forming these petals, however, at the resonance Fig.~\ref{figlin}(a) mode $m=4$ appears to have dominant contribution. } Under the flipping of the polarization from RCP to LCP (corresponding to sign change from $+$ to $-$) the angular behaviour of the field intensity changes.  However, the overall integral intensity does not depend on the signs $\pm$, and two types of modes contribute independently and equally for RCP and LCP: ${I}^{2\omega, total}=I_0+I_4$, somewhat resembling the mechanism of a hidden chirality discussed previously~\cite{Horrer2020-LocalOpticalChirali, Zu2018-Deep-SubwavelengthRe}.

\subsection{SHG for nanoparticle dimers} 

Now, we consider a dimer composed of two identical cylindrical nanoparticles. We assume that the polarisation $ P^{2\omega} (r, z, \varphi)$ remains the same in each single nanoparticle being similar to that of an isolated nanoparticle, and we study the effect of mode hybridisation on the SH fields. 


In our approximation, the difference between isolated nanoparticles and nanoparticle dimers appears at the level of the coupling integral. In case of a single nanoparticle, the modes proportional to $\sin(4\varphi)$ and $\cos(4\varphi)$ are degenerate. For a dimer, the hybridisation of these two degenerate modes leads to four modes with different symmetries and energies (see Fig.~\ref{fig3}(a) )\cite{Nordlander2004-PlasmonHybridization,Gao2017-ForwardBackwardSwit,Deng2018-DarkPlasmonModesin,Pascale2019-Full-waveelectromagn,PhysRevB.92.045433,Dmitriev2021-Opticalcouplingofo,Song2021-Nanoelectromechanical}. According to the selection rules~\cite{Frizyuk_2019}, only the modes $A_1$ and $A_2$ are excited in the SH field, {and comparing Fig.~\ref{figlin}(a) and Fig.~\ref{fig4}(a) we see how one resonant peak ($E_4$) transformed into two ($A_1$ and $A_2$)}. Based on the mode symmetry, we can calculate the coupling integral over one nanoparticle, while the integral over the second nanoparticle will be the same. 
First, we use the approximation that only single multipole $\propto \cos(4\varphi)$ dominates in each nanoparticle. However, the far-field radiation pattern has a complex multipole composition as a sum of fields generated by two nanoparticles as shown in the table of Fig.~\ref{fig3}(c).

\begin{widetext}

The coupling integral with this mode takes the form
\begin{figure}[ht!]
  \includegraphics[width=0.45\linewidth]{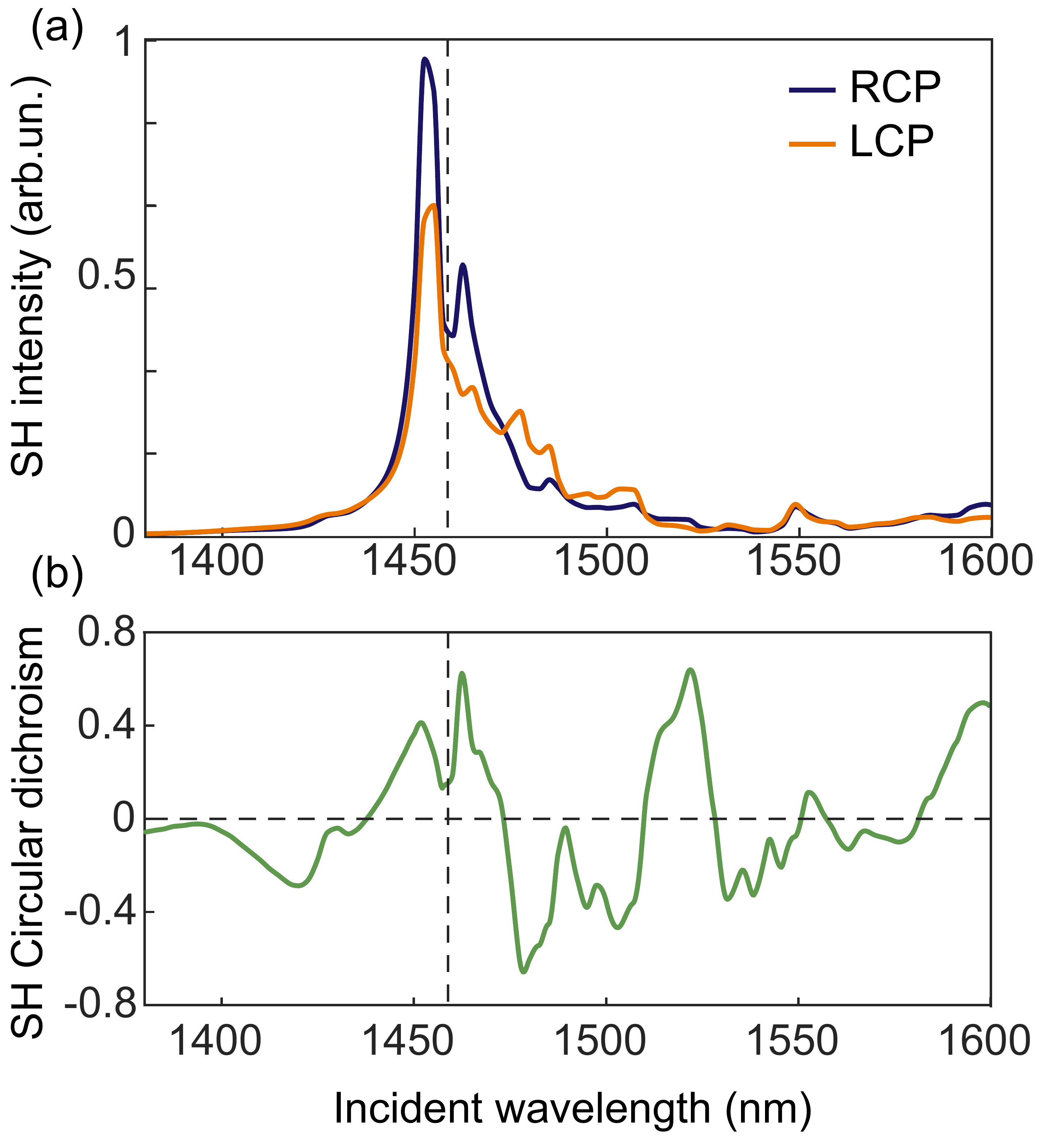}
\caption{Numerically calculated (a) SH intensity for both RCP and LCP fundamental beams, and (b) resulting nonlinear circular dichroism SH-CD=$2(I^{2\omega}_{RCP}-I^{2\omega}_{LCP})/(I^{2\omega}_{RCP}+I^{2\omega}_{LCP})$\cite{Belardini2020-CircularDichroismin,Schmeltz2020-Circulardichroismse} of the SH signal for the nanoparticle dimer. }\label{fig4}
\end{figure}
\begin{equation}
  D^{dim}_4\propto \int_0^{2\pi} d\varphi \cos(4\varphi)\left[1-e^{\pm 4i(\varphi-\beta)}\right]\propto e^{\mp 4i \beta}.
\end{equation}

Now, we calculate the coupling integral for the mode with dominating $m=0$ multipoles, 
\begin{equation}
  D^{dim}_0\propto \int_0^{2\pi} d\varphi (1-e^{\pm 4i(\varphi-\beta)})\propto 2\pi.
\end{equation}
\begin{figure}[ht!]
\includegraphics[width=0.75\linewidth]{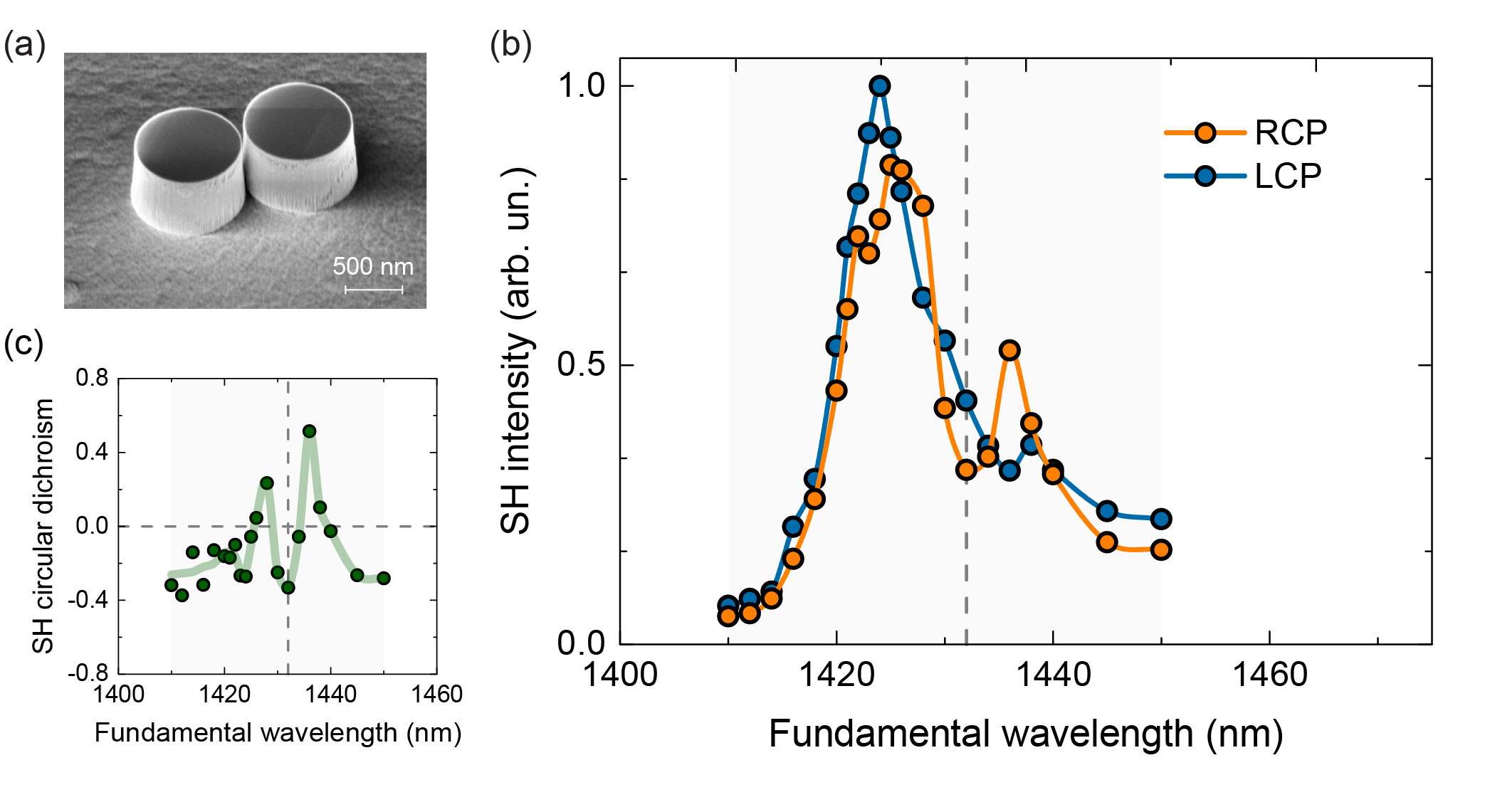}
\caption{Experimental SHG for a nanoparticle dimer. (a)~Scanning electron micrograph of the nanoparticle dimer with diameters 945~$\pm$~5~nm and height 635~nm. (b)~Measured SH spectra of the nanoparticle pumped by circularly polarized laser beams: RCP (orange) and LCP (blue). (c)~Circular dichroism obtained from the SH signals of (b) generated by the nanoparticle dimer. Thick solid line is a guide for an eye.}
\label{fig5}
\end{figure}
\end{widetext}

The most important result of this analysis is that these two modes can contribute to the overall intensity, while they both are transformed by the $A_1$  representation and radiate into the same radiation channel.  Being excited with different relative phases, the mode overlap results in strong dichroism. Indeed, let us consider a very illustrative case  when one of these two modes is resonant and the second is not. The phase difference between the mode coefficients equals $\pi/2$; multipolar contributions to $A_1$ SH far-field (Fig.~\ref{fig3}(c))  
will come with different phases $c_{\vec W}=a+\exp(i\pi/2)b \exp(\mp4i\beta)=a+ib \exp(\mp4i\beta)$, here $c_{\vec W}$ is the amplitude of a particular multipole. Now one can immediately see that this amplitude starts to have strong polarization dependence. When $\beta=Z\pi/4$, $Z=0,1,2..$, the difference between LCP and RCP beam excitations should not be observed in SH signal, {and that,  indeed, was confirmed in our  numerical simulations.} 
Once  $\beta=\pi/8$, for LCP we can write $a+ibe^{-i\pi/2}=a+b$, and for RCP  the same coefficient will become $a-b$,  and CD is maximal {in this case. We believe, that by proper engineering modes' phases, amplitudes, and $\beta$ one can achieve even higher CD values.} 
 For this particular $\beta$, we obtain a local maximum close to resonance (Fig.~\ref{fig4}(b)), however, the resonance is not required for the existence of CD. The required phases and amplitudes can appear for nonresonant wavelengths, as well as the conditions will be different for different values of $\beta$. For $A_2$ modes all considerations are analogous. 


In order to verify our general analysis and theoretical findings, we conduct nonlinear spectroscopy from the dimers of two AlGaAs nanoparticles fabricated by a consequent transfer of individual nanoparticles on a glass substrate as described earlier~\cite{melik2021}. The dimers are formed (see an SEM picture in Fig.~\ref{fig5}(a)) with the random relative orientation of crystalline lattices, which breaks the symmetry required for observation of CD in SHG. The dimer is illuminated normally by circularly polarized focused optical pulses of 2 ps duration in 1410-1450~nm range. The excitation occurs through an objective with 0.7~NA from above the substrate. The generated SH signal is collected by an objective lens with 0.9~NA. We observe experimentally the effect of nonlinear CD in the dimer, as shown in Fig.~\ref{fig5}(b). More specifically, the nonlinear CD is enhanced at the resonant fundamental frequency, and, at the same time, it takes place at nonresonant frequencies (see Fig.~\ref{fig5}(c)) in accordance with the theoretical predictions. The difference between the simulated spectra (Fig.~\ref{fig4}) and experimental results can be attributed to a different geometry of the cylindrical nanoparticles being  slightly conical and not absolutely identical in the experiment. We note that the relative angle of the crystalline axis in two nanoparticles is arbitrary, and it is hard to determine it exactly in our case. Consequently, these experimental results reflect well the major predictions of our general multipolar hybridization theory, and also confirm a dramatic enhancement of all the effects in the vicinity of Mie resonances. 

In summary, we have systematically studied the SHG from the dimer composed  of two identical III-V semiconductor nanoparticles. We have illuminated the dimer  by a circularly polarized light and observed, both theoretically and experimentally,  that the nonlinear signal possesses circular dichroism. We have revealed that this effect can be attributed to the Mie-resonant multipolar response of the nanoparticles, and demonstrated that the result is nonzero when the dimer axes is oriented under an angle to the crystalline lattice of the dielectric material. 

\section{Acknowledgement}
The authors thank S.~Kruk for a technical assistance, as well as P. Banzer, A. Krasnok, and V.~Valev for useful comments and suggestions. K.F. thanks A.A.~Bogdanov, K.~Koshelev  and A.A.~Nikolaeva for  fruitful discussions. This work was supported by the Australian Research Council (grant DP210101292), the National Research Foundation (NRF) of Korea (grant no. 2021R1A2C3006781 funded by the Korean Government, MSIT) and the Russian Science Foundation (grant no. 18-72-10140). E.M. acknowledges a support of the ACT Node of the Australian National Fabrication Facility. H.-G.P. acknowledges a support from the Samsung Research Funding \& Incubation Center of Samsung Electronics (SRFC-MA2001-01). K.F. acknowledges support
from the Foundation for the Advancement of Theoretical
Physics and Mathematics "BASIS" (Russia).
\bibliography{bibfile}

\begin{thebibliography}{68}%
\makeatletter
\providecommand \@ifxundefined [1]{%
 \@ifx{#1\undefined}
}%
\providecommand \@ifnum [1]{%
 \ifnum #1\expandafter \@firstoftwo
 \else \expandafter \@secondoftwo
 \fi
}%
\providecommand \@ifx [1]{%
 \ifx #1\expandafter \@firstoftwo
 \else \expandafter \@secondoftwo
 \fi
}%
\providecommand \natexlab [1]{#1}%
\providecommand \enquote  [1]{``#1''}%
\providecommand \bibnamefont  [1]{#1}%
\providecommand \bibfnamefont [1]{#1}%
\providecommand \citenamefont [1]{#1}%
\providecommand \href@noop [0]{\@secondoftwo}%
\providecommand \href [0]{\begingroup \@sanitize@url \@href}%
\providecommand \@href[1]{\@@startlink{#1}\@@href}%
\providecommand \@@href[1]{\endgroup#1\@@endlink}%
\providecommand \@sanitize@url [0]{\catcode `\\12\catcode `\$12\catcode
  `\&12\catcode `\#12\catcode `\^12\catcode `\_12\catcode `\%12\relax}%
\providecommand \@@startlink[1]{}%
\providecommand \@@endlink[0]{}%
\providecommand \url  [0]{\begingroup\@sanitize@url \@url }%
\providecommand \@url [1]{\endgroup\@href {#1}{\urlprefix }}%
\providecommand \urlprefix  [0]{URL }%
\providecommand \Eprint [0]{\href }%
\providecommand \doibase [0]{https://doi.org/}%
\providecommand \selectlanguage [0]{\@gobble}%
\providecommand \bibinfo  [0]{\@secondoftwo}%
\providecommand \bibfield  [0]{\@secondoftwo}%
\providecommand \translation [1]{[#1]}%
\providecommand \BibitemOpen [0]{}%
\providecommand \bibitemStop [0]{}%
\providecommand \bibitemNoStop [0]{.\EOS\space}%
\providecommand \EOS [0]{\spacefactor3000\relax}%
\providecommand \BibitemShut  [1]{\csname bibitem#1\endcsname}%
\let\auto@bib@innerbib\@empty
\bibitem [{\citenamefont {Berova}\ \emph {et~al.}(2000)\citenamefont {Berova},
  \citenamefont {Nakanishi},\ and\ \citenamefont {Woody}}]{Nakanishi1995}%
  \BibitemOpen
  \bibfield  {author} {\bibinfo {author} {\bibfnamefont {N.}~\bibnamefont
  {Berova}}, \bibinfo {author} {\bibfnamefont {K.}~\bibnamefont {Nakanishi}},\
  and\ \bibinfo {author} {\bibfnamefont {R.~W.}\ \bibnamefont {Woody}},\ }\href
  {https://www.wiley.com/en-us/Circular+Dichroism\%3A+Principles+and+Applications\%2C+2nd+Edition-p-9780471330035}
  {\emph {\bibinfo {title} {{Circular Dichroism: Principles and Applications,
  2nd Edition}}}}\ (\bibinfo  {publisher} {Wiley-VCH},\ \bibinfo {address}
  {Hoboken, NJ, USA},\ \bibinfo {year} {2000})\BibitemShut {NoStop}%
\bibitem [{\citenamefont {Greenfield}(2006)}]{Greenfield_2006}%
  \BibitemOpen
  \bibfield  {author} {\bibinfo {author} {\bibfnamefont {N.~J.}\ \bibnamefont
  {Greenfield}},\ }\bibfield  {title} {\bibinfo {title} {Using circular
  dichroism spectra to estimate protein secondary structure},\ }\href
  {https://doi.org/10.1038/nprot.2006.202} {\bibfield  {journal} {\bibinfo
  {journal} {Nature Protocols}\ }\textbf {\bibinfo {volume} {1}},\ \bibinfo
  {pages} {2876} (\bibinfo {year} {2006})}\BibitemShut {NoStop}%
\bibitem [{\citenamefont {Hopkins}\ \emph {et~al.}(2016)\citenamefont
  {Hopkins}, \citenamefont {Poddubny}, \citenamefont {Miroshnichenko},\ and\
  \citenamefont {Kivshar}}]{Hopkins2016-Circulardichro}%
  \BibitemOpen
  \bibfield  {author} {\bibinfo {author} {\bibfnamefont {B.}~\bibnamefont
  {Hopkins}}, \bibinfo {author} {\bibfnamefont {A.~N.}\ \bibnamefont
  {Poddubny}}, \bibinfo {author} {\bibfnamefont {A.~E.}\ \bibnamefont
  {Miroshnichenko}},\ and\ \bibinfo {author} {\bibfnamefont {Y.~S.}\
  \bibnamefont {Kivshar}},\ }\bibfield  {title} {\bibinfo {title} {{Circular
  dichroism induced by Fano resonances in planar chiral oligomers}},\ }\href
  {https://doi.org/10.1002/lpor.201500222} {\bibfield  {journal} {\bibinfo
  {journal} {Laser Photonics Rev.}\ }\textbf {\bibinfo {volume} {10}},\
  \bibinfo {pages} {137} (\bibinfo {year} {2016})}\BibitemShut {NoStop}%
\bibitem [{\citenamefont {De~Silva}\ \emph {et~al.}(2021)\citenamefont
  {De~Silva}, \citenamefont {Atri-Schuller}, \citenamefont {Dubey},
  \citenamefont {Acharya}, \citenamefont {Romans}, \citenamefont {Foster},
  \citenamefont {Russ}, \citenamefont {Compton}, \citenamefont {Rischbieter},
  \citenamefont {Douguet}, \citenamefont {Bartschat},\ and\ \citenamefont
  {Fischer}}]{DeSilva2021-UsingCircularDichro}%
  \BibitemOpen
  \bibfield  {author} {\bibinfo {author} {\bibfnamefont {A.~H. N.~C.}\
  \bibnamefont {De~Silva}}, \bibinfo {author} {\bibfnamefont {D.}~\bibnamefont
  {Atri-Schuller}}, \bibinfo {author} {\bibfnamefont {S.}~\bibnamefont
  {Dubey}}, \bibinfo {author} {\bibfnamefont {B.~P.}\ \bibnamefont {Acharya}},
  \bibinfo {author} {\bibfnamefont {K.~L.}\ \bibnamefont {Romans}}, \bibinfo
  {author} {\bibfnamefont {K.}~\bibnamefont {Foster}}, \bibinfo {author}
  {\bibfnamefont {O.}~\bibnamefont {Russ}}, \bibinfo {author} {\bibfnamefont
  {K.}~\bibnamefont {Compton}}, \bibinfo {author} {\bibfnamefont
  {C.}~\bibnamefont {Rischbieter}}, \bibinfo {author} {\bibfnamefont
  {N.}~\bibnamefont {Douguet}}, \bibinfo {author} {\bibfnamefont
  {K.}~\bibnamefont {Bartschat}},\ and\ \bibinfo {author} {\bibfnamefont
  {D.}~\bibnamefont {Fischer}},\ }\bibfield  {title} {\bibinfo {title} {{Using
  Circular Dichroism to Control Energy Transfer in Multiphoton Ionization}},\
  }\href {https://doi.org/10.1103/PhysRevLett.126.023201} {\bibfield  {journal}
  {\bibinfo  {journal} {Phys. Rev. Lett.}\ }\textbf {\bibinfo {volume} {126}},\
  \bibinfo {pages} {023201} (\bibinfo {year} {2021})}\BibitemShut {NoStop}%
\bibitem [{\citenamefont {Sch\"aferling}\ \emph {et~al.}(2012)\citenamefont
  {Sch\"aferling}, \citenamefont {Dregely}, \citenamefont {Hentschel},\ and\
  \citenamefont {Giessen}}]{PhysRevX.2.031010}%
  \BibitemOpen
  \bibfield  {author} {\bibinfo {author} {\bibfnamefont {M.}~\bibnamefont
  {Sch\"aferling}}, \bibinfo {author} {\bibfnamefont {D.}~\bibnamefont
  {Dregely}}, \bibinfo {author} {\bibfnamefont {M.}~\bibnamefont {Hentschel}},\
  and\ \bibinfo {author} {\bibfnamefont {H.}~\bibnamefont {Giessen}},\
  }\bibfield  {title} {\bibinfo {title} {Tailoring enhanced optical chirality:
  Design principles for chiral plasmonic anostructures},\ }\href
  {https://doi.org/10.1103/PhysRevX.2.031010} {\bibfield  {journal} {\bibinfo
  {journal} {Phys. Rev. X}\ }\textbf {\bibinfo {volume} {2}},\ \bibinfo {pages}
  {031010} (\bibinfo {year} {2012})}\BibitemShut {NoStop}%
\bibitem [{\citenamefont {Hentschel}\ \emph {et~al.}(2017)\citenamefont
  {Hentschel}, \citenamefont {Sch\"aferling}, \citenamefont {Duan},
  \citenamefont {Giessen},\ and\ \citenamefont {Liu}}]{Na_Liu2017}%
  \BibitemOpen
  \bibfield  {author} {\bibinfo {author} {\bibfnamefont {M.}~\bibnamefont
  {Hentschel}}, \bibinfo {author} {\bibfnamefont {M.}~\bibnamefont
  {Sch\"aferling}}, \bibinfo {author} {\bibfnamefont {X.}~\bibnamefont {Duan}},
  \bibinfo {author} {\bibfnamefont {H.}~\bibnamefont {Giessen}},\ and\ \bibinfo
  {author} {\bibfnamefont {N.}~\bibnamefont {Liu}},\ }\bibfield  {title}
  {\bibinfo {title} {Chiral plasmonics},\ }\href@noop {} {\bibfield  {journal}
  {\bibinfo  {journal} {Science Advances}\ }\textbf {\bibinfo {volume} {3}},\
  \bibinfo {pages} {e1602737} (\bibinfo {year} {2017})}\BibitemShut {NoStop}%
\bibitem [{\citenamefont {Collins}\ \emph {et~al.}(2017)\citenamefont
  {Collins}, \citenamefont {Kuppe}, \citenamefont {Hooper}, \citenamefont
  {Sibilia}, \citenamefont {Centini},\ and\ \citenamefont {Valev}}]{Valev2017}%
  \BibitemOpen
  \bibfield  {author} {\bibinfo {author} {\bibfnamefont {J.}~\bibnamefont
  {Collins}}, \bibinfo {author} {\bibfnamefont {C.}~\bibnamefont {Kuppe}},
  \bibinfo {author} {\bibfnamefont {D.}~\bibnamefont {Hooper}}, \bibinfo
  {author} {\bibfnamefont {C.}~\bibnamefont {Sibilia}}, \bibinfo {author}
  {\bibfnamefont {M.}~\bibnamefont {Centini}},\ and\ \bibinfo {author}
  {\bibfnamefont {V.}~\bibnamefont {Valev}},\ }\bibfield  {title} {\bibinfo
  {title} {Chirality and chiroptical effects in metal nanostructures:
  fundamentals and current trends},\ }\href@noop {} {\bibfield  {journal}
  {\bibinfo  {journal} {Advanced Optical Materials}\ }\textbf {\bibinfo
  {volume} {5}},\ \bibinfo {pages} {1700182} (\bibinfo {year}
  {2017})}\BibitemShut {NoStop}%
\bibitem [{\citenamefont {Li}\ \emph {et~al.}(2021)\citenamefont {Li},
  \citenamefont {Wang}, \citenamefont {Wu}, \citenamefont {Li}, \citenamefont
  {Hu}, \citenamefont {Jiang}, \citenamefont {Guo}, \citenamefont {Liu},
  \citenamefont {Yao}, \citenamefont {Chen}, \citenamefont {Fang},
  \citenamefont {Fan}, \citenamefont {Korgel}, \citenamefont
  {Al{\ifmmode\grave{u}\else\`{u}\fi}},\ and\ \citenamefont
  {Zheng}}]{Li2021-TunableChiralOptics}%
  \BibitemOpen
  \bibfield  {author} {\bibinfo {author} {\bibfnamefont {J.}~\bibnamefont
  {Li}}, \bibinfo {author} {\bibfnamefont {M.}~\bibnamefont {Wang}}, \bibinfo
  {author} {\bibfnamefont {Z.}~\bibnamefont {Wu}}, \bibinfo {author}
  {\bibfnamefont {H.}~\bibnamefont {Li}}, \bibinfo {author} {\bibfnamefont
  {G.}~\bibnamefont {Hu}}, \bibinfo {author} {\bibfnamefont {T.}~\bibnamefont
  {Jiang}}, \bibinfo {author} {\bibfnamefont {J.}~\bibnamefont {Guo}}, \bibinfo
  {author} {\bibfnamefont {Y.}~\bibnamefont {Liu}}, \bibinfo {author}
  {\bibfnamefont {K.}~\bibnamefont {Yao}}, \bibinfo {author} {\bibfnamefont
  {Z.}~\bibnamefont {Chen}}, \bibinfo {author} {\bibfnamefont {J.}~\bibnamefont
  {Fang}}, \bibinfo {author} {\bibfnamefont {D.}~\bibnamefont {Fan}}, \bibinfo
  {author} {\bibfnamefont {B.~A.}\ \bibnamefont {Korgel}}, \bibinfo {author}
  {\bibfnamefont {A.}~\bibnamefont {Al{\ifmmode\grave{u}\else\`{u}\fi}}},\ and\
  \bibinfo {author} {\bibfnamefont {Y.}~\bibnamefont {Zheng}},\ }\bibfield
  {title} {\bibinfo {title} {{Tunable Chiral Optics in All-Solid-Phase
  Reconfigurable Dielectric Nanostructures}},\ }\href
  {https://doi.org/10.1021/acs.nanolett.0c03957} {\bibfield  {journal}
  {\bibinfo  {journal} {Nano Lett.}\ }\textbf {\bibinfo {volume} {21}},\
  \bibinfo {pages} {973} (\bibinfo {year} {2021})}\BibitemShut {NoStop}%
\bibitem [{\citenamefont {Decker}\ \emph {et~al.}(2007)\citenamefont {Decker},
  \citenamefont {Klein}, \citenamefont {Wegener},\ and\ \citenamefont
  {Linden}}]{Decker_Klein_Wegener_Linden_2007}%
  \BibitemOpen
  \bibfield  {author} {\bibinfo {author} {\bibfnamefont {M.}~\bibnamefont
  {Decker}}, \bibinfo {author} {\bibfnamefont {M.~W.}\ \bibnamefont {Klein}},
  \bibinfo {author} {\bibfnamefont {M.}~\bibnamefont {Wegener}},\ and\ \bibinfo
  {author} {\bibfnamefont {S.}~\bibnamefont {Linden}},\ }\bibfield  {title}
  {\bibinfo {title} {Circular dichroism of planar chiral magnetic
  metamaterials},\ }\href {https://doi.org/10.1364/OL.32.000856} {\bibfield
  {journal} {\bibinfo  {journal} {Optics Letters}\ }\textbf {\bibinfo {volume}
  {32}},\ \bibinfo {pages} {856} (\bibinfo {year} {2007})}\BibitemShut
  {NoStop}%
\bibitem [{\citenamefont {Wang}\ \emph {et~al.}(2016)\citenamefont {Wang},
  \citenamefont {Cheng}, \citenamefont {Winsor},\ and\ \citenamefont
  {Liu}}]{Wang_Cheng_Winsor_Liu_2016}%
  \BibitemOpen
  \bibfield  {author} {\bibinfo {author} {\bibfnamefont {Z.}~\bibnamefont
  {Wang}}, \bibinfo {author} {\bibfnamefont {F.}~\bibnamefont {Cheng}},
  \bibinfo {author} {\bibfnamefont {T.}~\bibnamefont {Winsor}},\ and\ \bibinfo
  {author} {\bibfnamefont {Y.}~\bibnamefont {Liu}},\ }\bibfield  {title}
  {\bibinfo {title} {Optical chiral metamaterials: a review of the
  fundamentals, fabrication methods and applications},\ }\href
  {https://doi.org/10.1088/0957-4484/27/41/412001} {\bibfield  {journal}
  {\bibinfo  {journal} {Nanotechnology}\ }\textbf {\bibinfo {volume} {27}},\
  \bibinfo {pages} {412001} (\bibinfo {year} {2016})}\BibitemShut {NoStop}%
\bibitem [{\citenamefont {Zhao}\ \emph {et~al.}(2012)\citenamefont {Zhao},
  \citenamefont {Belkin},\ and\ \citenamefont
  {Al{\ifmmode\grave{u}\else\`{u}\fi}}}]{Zhao2012-Twistedopticalmetam}%
  \BibitemOpen
  \bibfield  {author} {\bibinfo {author} {\bibfnamefont {Y.}~\bibnamefont
  {Zhao}}, \bibinfo {author} {\bibfnamefont {M.~A.}\ \bibnamefont {Belkin}},\
  and\ \bibinfo {author} {\bibfnamefont {A.}~\bibnamefont
  {Al{\ifmmode\grave{u}\else\`{u}\fi}}},\ }\bibfield  {title} {\bibinfo {title}
  {{Twisted optical metamaterials for planarized ultrathin broadband circular
  polarizers}},\ }\href {https://doi.org/10.1038/ncomms1877} {\bibfield
  {journal} {\bibinfo  {journal} {Nat. Commun.}\ }\textbf {\bibinfo {volume}
  {3}},\ \bibinfo {pages} {1} (\bibinfo {year} {2012})}\BibitemShut {NoStop}%
\bibitem [{\citenamefont {Hu}\ \emph {et~al.}(2020)\citenamefont {Hu},
  \citenamefont {Lawrence},\ and\ \citenamefont {Dionne}}]{dionne}%
  \BibitemOpen
  \bibfield  {author} {\bibinfo {author} {\bibfnamefont {J.}~\bibnamefont
  {Hu}}, \bibinfo {author} {\bibfnamefont {M.}~\bibnamefont {Lawrence}},\ and\
  \bibinfo {author} {\bibfnamefont {J.~A.}\ \bibnamefont {Dionne}},\ }\bibfield
   {title} {\bibinfo {title} {{High Quality Factor Dielectric Metasurfaces for
  Ultraviolet Circular Dichroism Spectroscopy}},\ }\href
  {https://doi.org/10.1021/acsphotonics.9b01352} {\bibfield  {journal}
  {\bibinfo  {journal} {ACS Photonics}\ }\textbf {\bibinfo {volume} {7}},\
  \bibinfo {pages} {36} (\bibinfo {year} {2020})}\BibitemShut {NoStop}%
\bibitem [{\citenamefont {Gorkunov}\ \emph {et~al.}(2020)\citenamefont
  {Gorkunov}, \citenamefont {Antonov},\ and\ \citenamefont {Kivshar}}]{maxim}%
  \BibitemOpen
  \bibfield  {author} {\bibinfo {author} {\bibfnamefont {M.~V.}\ \bibnamefont
  {Gorkunov}}, \bibinfo {author} {\bibfnamefont {A.~A.}\ \bibnamefont
  {Antonov}},\ and\ \bibinfo {author} {\bibfnamefont {Y.~S.}\ \bibnamefont
  {Kivshar}},\ }\bibfield  {title} {\bibinfo {title} {{Metasurfaces with
  Maximum Chirality Empowered by Bound States in the Continuum}},\ }\href
  {https://doi.org/10.1103/PhysRevLett.125.093903} {\bibfield  {journal}
  {\bibinfo  {journal} {Phys. Rev. Lett.}\ }\textbf {\bibinfo {volume} {125}},\
  \bibinfo {pages} {093903} (\bibinfo {year} {2020})}\BibitemShut {NoStop}%
\bibitem [{\citenamefont {Chen}\ \emph {et~al.}(2021)\citenamefont {Chen},
  \citenamefont {Gao}, \citenamefont {Song}, \citenamefont {Li}, \citenamefont
  {Zhu},\ and\ \citenamefont {Li}}]{tao_li}%
  \BibitemOpen
  \bibfield  {author} {\bibinfo {author} {\bibfnamefont {C.}~\bibnamefont
  {Chen}}, \bibinfo {author} {\bibfnamefont {S.}~\bibnamefont {Gao}}, \bibinfo
  {author} {\bibfnamefont {W.}~\bibnamefont {Song}}, \bibinfo {author}
  {\bibfnamefont {H.}~\bibnamefont {Li}}, \bibinfo {author} {\bibfnamefont
  {S.-N.}\ \bibnamefont {Zhu}},\ and\ \bibinfo {author} {\bibfnamefont
  {T.}~\bibnamefont {Li}},\ }\bibfield  {title} {\bibinfo {title} {Metasurfaces
  with planar chiral meta-atoms for spin light manipulation},\ }\href
  {https://doi.org/10.1021/acs.nanolett.0c04902} {\bibfield  {journal}
  {\bibinfo  {journal} {Nano Letters}\ }\textbf {\bibinfo {volume} {21}},\
  \bibinfo {pages} {XXXX} (\bibinfo {year} {2021})}\BibitemShut {NoStop}%
\bibitem [{\citenamefont {Lin}\ \emph {et~al.}(2019)\citenamefont {Lin},
  \citenamefont {Lepeshov}, \citenamefont {Krasnok}, \citenamefont {Jiang},
  \citenamefont {Peng}, \citenamefont {Korgel}, \citenamefont
  {Al{\ifmmode\grave{u}\else\`{u}\fi}},\ and\ \citenamefont
  {Zheng}}]{Krasnok2019}%
  \BibitemOpen
  \bibfield  {author} {\bibinfo {author} {\bibfnamefont {L.}~\bibnamefont
  {Lin}}, \bibinfo {author} {\bibfnamefont {S.}~\bibnamefont {Lepeshov}},
  \bibinfo {author} {\bibfnamefont {A.}~\bibnamefont {Krasnok}}, \bibinfo
  {author} {\bibfnamefont {T.}~\bibnamefont {Jiang}}, \bibinfo {author}
  {\bibfnamefont {X.}~\bibnamefont {Peng}}, \bibinfo {author} {\bibfnamefont
  {B.~A.}\ \bibnamefont {Korgel}}, \bibinfo {author} {\bibfnamefont
  {A.}~\bibnamefont {Al{\ifmmode\grave{u}\else\`{u}\fi}}},\ and\ \bibinfo
  {author} {\bibfnamefont {Y.}~\bibnamefont {Zheng}},\ }\bibfield  {title}
  {\bibinfo {title} {{All-optical reconfigurable chiral meta-molecules}},\
  }\href {https://doi.org/10.1016/j.mattod.2019.02.015} {\bibfield  {journal}
  {\bibinfo  {journal} {Mater. Today}\ }\textbf {\bibinfo {volume} {25}},\
  \bibinfo {pages} {10} (\bibinfo {year} {2019})}\BibitemShut {NoStop}%
\bibitem [{\citenamefont {Nechayev}\ \emph {et~al.}(2019)\citenamefont
  {Nechayev}, \citenamefont {Barczyk}, \citenamefont {Mick},\ and\
  \citenamefont {Banzer}}]{Banzer2019}%
  \BibitemOpen
  \bibfield  {author} {\bibinfo {author} {\bibfnamefont {S.}~\bibnamefont
  {Nechayev}}, \bibinfo {author} {\bibfnamefont {R.}~\bibnamefont {Barczyk}},
  \bibinfo {author} {\bibfnamefont {U.}~\bibnamefont {Mick}},\ and\ \bibinfo
  {author} {\bibfnamefont {P.}~\bibnamefont {Banzer}},\ }\bibfield  {title}
  {\bibinfo {title} {Substrate-induced chirality in an individual
  nanostructure},\ }\href@noop {} {\bibfield  {journal} {\bibinfo  {journal}
  {ACS Photonics}\ }\textbf {\bibinfo {volume} {6}},\ \bibinfo {pages} {1876}
  (\bibinfo {year} {2019})}\BibitemShut {NoStop}%
\bibitem [{\citenamefont {Nechayev}\ \emph {et~al.}(2018)\citenamefont
  {Nechayev}, \citenamefont {Wo{\ifmmode\acute{z}\else\'{z}\fi}niak},
  \citenamefont {Neugebauer}, \citenamefont {Barczyk},\ and\ \citenamefont
  {Banzer}}]{Nechayev2018-ChiralityofSymmetr}%
  \BibitemOpen
  \bibfield  {author} {\bibinfo {author} {\bibfnamefont {S.}~\bibnamefont
  {Nechayev}}, \bibinfo {author} {\bibfnamefont {P.}~\bibnamefont
  {Wo{\ifmmode\acute{z}\else\'{z}\fi}niak}}, \bibinfo {author} {\bibfnamefont
  {M.}~\bibnamefont {Neugebauer}}, \bibinfo {author} {\bibfnamefont
  {R.}~\bibnamefont {Barczyk}},\ and\ \bibinfo {author} {\bibfnamefont
  {P.}~\bibnamefont {Banzer}},\ }\bibfield  {title} {\bibinfo {title}
  {{Chirality of Symmetric Resonant Heterostructures}},\ }\href
  {https://doi.org/10.1002/lpor.201800109} {\bibfield  {journal} {\bibinfo
  {journal} {Laser Photonics Rev.}\ }\textbf {\bibinfo {volume} {12}},\
  \bibinfo {pages} {1800109} (\bibinfo {year} {2018})}\BibitemShut {NoStop}%
\bibitem [{\citenamefont {Bautista}\ \emph {et~al.}(2012)\citenamefont
  {Bautista}, \citenamefont {Huttunen}, \citenamefont
  {M{\ifmmode\ddot{a}\else\"{a}\fi}kitalo}, \citenamefont {Kontio},
  \citenamefont {Simonen},\ and\ \citenamefont
  {Kauranen}}]{Bautista2012-Second-HarmonicGener}%
  \BibitemOpen
  \bibfield  {author} {\bibinfo {author} {\bibfnamefont {G.}~\bibnamefont
  {Bautista}}, \bibinfo {author} {\bibfnamefont {M.~J.}\ \bibnamefont
  {Huttunen}}, \bibinfo {author} {\bibfnamefont {J.}~\bibnamefont
  {M{\ifmmode\ddot{a}\else\"{a}\fi}kitalo}}, \bibinfo {author} {\bibfnamefont
  {J.~M.}\ \bibnamefont {Kontio}}, \bibinfo {author} {\bibfnamefont
  {J.}~\bibnamefont {Simonen}},\ and\ \bibinfo {author} {\bibfnamefont
  {M.}~\bibnamefont {Kauranen}},\ }\bibfield  {title} {\bibinfo {title}
  {{Second-Harmonic Generation Imaging of Metal Nano-Objects with Cylindrical
  Vector Beams}},\ }\href {https://doi.org/10.1021/nl301190x} {\bibfield
  {journal} {\bibinfo  {journal} {Nano Lett.}\ }\textbf {\bibinfo {volume}
  {12}},\ \bibinfo {pages} {3207} (\bibinfo {year} {2012})}\BibitemShut
  {NoStop}%
\bibitem [{\citenamefont {Belardini}\ \emph {et~al.}(2011)\citenamefont
  {Belardini}, \citenamefont {Larciprete}, \citenamefont {Centini},
  \citenamefont {Fazio}, \citenamefont {Sibilia}, \citenamefont {Chiappe},
  \citenamefont {Martella}, \citenamefont {Toma}, \citenamefont {Giordano},\
  and\ \citenamefont {Buatier~de Mongeot}}]{PhysRevLett.107.257401}%
  \BibitemOpen
  \bibfield  {author} {\bibinfo {author} {\bibfnamefont {A.}~\bibnamefont
  {Belardini}}, \bibinfo {author} {\bibfnamefont {M.~C.}\ \bibnamefont
  {Larciprete}}, \bibinfo {author} {\bibfnamefont {M.}~\bibnamefont {Centini}},
  \bibinfo {author} {\bibfnamefont {E.}~\bibnamefont {Fazio}}, \bibinfo
  {author} {\bibfnamefont {C.}~\bibnamefont {Sibilia}}, \bibinfo {author}
  {\bibfnamefont {D.}~\bibnamefont {Chiappe}}, \bibinfo {author} {\bibfnamefont
  {C.}~\bibnamefont {Martella}}, \bibinfo {author} {\bibfnamefont
  {A.}~\bibnamefont {Toma}}, \bibinfo {author} {\bibfnamefont {M.}~\bibnamefont
  {Giordano}},\ and\ \bibinfo {author} {\bibfnamefont {F.}~\bibnamefont
  {Buatier~de Mongeot}},\ }\bibfield  {title} {\bibinfo {title} {Circular
  dichroism in the optical second-harmonic emission of curved gold metal
  nanowires},\ }\href {https://doi.org/10.1103/PhysRevLett.107.257401}
  {\bibfield  {journal} {\bibinfo  {journal} {Phys. Rev. Lett.}\ }\textbf
  {\bibinfo {volume} {107}},\ \bibinfo {pages} {257401} (\bibinfo {year}
  {2011})}\BibitemShut {NoStop}%
\bibitem [{\citenamefont {Hooper}\ \emph {et~al.}(2017)\citenamefont {Hooper},
  \citenamefont {Mark}, \citenamefont {Kuppe}, \citenamefont {Collins},
  \citenamefont {Fischer},\ and\ \citenamefont
  {Valev}}]{Hooper_Mark_Kuppe_Collins_Fischer_Valev_2017}%
  \BibitemOpen
  \bibfield  {author} {\bibinfo {author} {\bibfnamefont {D.~C.}\ \bibnamefont
  {Hooper}}, \bibinfo {author} {\bibfnamefont {A.~G.}\ \bibnamefont {Mark}},
  \bibinfo {author} {\bibfnamefont {C.}~\bibnamefont {Kuppe}}, \bibinfo
  {author} {\bibfnamefont {J.~T.}\ \bibnamefont {Collins}}, \bibinfo {author}
  {\bibfnamefont {P.}~\bibnamefont {Fischer}},\ and\ \bibinfo {author}
  {\bibfnamefont {V.~K.}\ \bibnamefont {Valev}},\ }\bibfield  {title} {\bibinfo
  {title} {Strong rotational anisotropies affect nonlinear chiral
  metamaterials},\ }\href
  {https://doi.org/https://doi.org/10.1002/adma.201605110} {\bibfield
  {journal} {\bibinfo  {journal} {Advanced Materials}\ }\textbf {\bibinfo
  {volume} {29}},\ \bibinfo {pages} {1605110} (\bibinfo {year}
  {2017})}\BibitemShut {NoStop}%
\bibitem [{\citenamefont {Belardini}\ \emph {et~al.}(2020)\citenamefont
  {Belardini}, \citenamefont {Leahu}, \citenamefont {Petronijevic},
  \citenamefont {Hakkarainen}, \citenamefont {Koivusalo}, \citenamefont
  {Rizzo~Piton}, \citenamefont {Talmila}, \citenamefont {Guina},\ and\
  \citenamefont {Sibilia}}]{Belardini2020-CircularDichroismin}%
  \BibitemOpen
  \bibfield  {author} {\bibinfo {author} {\bibfnamefont {A.}~\bibnamefont
  {Belardini}}, \bibinfo {author} {\bibfnamefont {G.}~\bibnamefont {Leahu}},
  \bibinfo {author} {\bibfnamefont {E.}~\bibnamefont {Petronijevic}}, \bibinfo
  {author} {\bibfnamefont {T.}~\bibnamefont {Hakkarainen}}, \bibinfo {author}
  {\bibfnamefont {E.}~\bibnamefont {Koivusalo}}, \bibinfo {author}
  {\bibfnamefont {M.}~\bibnamefont {Rizzo~Piton}}, \bibinfo {author}
  {\bibfnamefont {S.}~\bibnamefont {Talmila}}, \bibinfo {author} {\bibfnamefont
  {M.}~\bibnamefont {Guina}},\ and\ \bibinfo {author} {\bibfnamefont
  {C.}~\bibnamefont {Sibilia}},\ }\bibfield  {title} {\bibinfo {title}
  {{Circular Dichroism in the Second Harmonic Field Evidenced by Asymmetric Au
  Coated GaAs Nanowires}},\ }\href {https://doi.org/10.3390/mi11020225}
  {\bibfield  {journal} {\bibinfo  {journal} {Micromachines}\ }\textbf
  {\bibinfo {volume} {11}},\ \bibinfo {pages} {225} (\bibinfo {year}
  {2020})}\BibitemShut {NoStop}%
\bibitem [{\citenamefont {Rodrigues}\ \emph {et~al.}(2014)\citenamefont
  {Rodrigues}, \citenamefont {Lan}, \citenamefont {Kang}, \citenamefont {Cui},\
  and\ \citenamefont {Cai}}]{Rodrigues2014-NonlinearImagingand}%
  \BibitemOpen
  \bibfield  {author} {\bibinfo {author} {\bibfnamefont {S.~P.}\ \bibnamefont
  {Rodrigues}}, \bibinfo {author} {\bibfnamefont {S.}~\bibnamefont {Lan}},
  \bibinfo {author} {\bibfnamefont {L.}~\bibnamefont {Kang}}, \bibinfo {author}
  {\bibfnamefont {Y.}~\bibnamefont {Cui}},\ and\ \bibinfo {author}
  {\bibfnamefont {W.}~\bibnamefont {Cai}},\ }\bibfield  {title} {\bibinfo
  {title} {{Nonlinear Imaging and Spectroscopy of Chiral Metamaterials}},\
  }\href {https://doi.org/10.1002/adma.201402293} {\bibfield  {journal}
  {\bibinfo  {journal} {Adv. Mater.}\ }\textbf {\bibinfo {volume} {26}},\
  \bibinfo {pages} {6157} (\bibinfo {year} {2014})}\BibitemShut {NoStop}%
\bibitem [{\citenamefont {Bertolotti}\ \emph {et~al.}(2015)\citenamefont
  {Bertolotti}, \citenamefont {Belardini}, \citenamefont {Benedetti},\ and\
  \citenamefont {Sibilia}}]{Bertolotti2015-Secondharmoniccircu}%
  \BibitemOpen
  \bibfield  {author} {\bibinfo {author} {\bibfnamefont {M.}~\bibnamefont
  {Bertolotti}}, \bibinfo {author} {\bibfnamefont {A.}~\bibnamefont
  {Belardini}}, \bibinfo {author} {\bibfnamefont {A.}~\bibnamefont
  {Benedetti}},\ and\ \bibinfo {author} {\bibfnamefont {C.}~\bibnamefont
  {Sibilia}},\ }\bibfield  {title} {\bibinfo {title} {{Second harmonic circular
  dichroism by self-assembled metasurfaces [Invited]}},\ }\href
  {https://doi.org/10.1364/JOSAB.32.001287} {\bibfield  {journal} {\bibinfo
  {journal} {J. Opt. Soc. Am. B, JOSAB}\ }\textbf {\bibinfo {volume} {32}},\
  \bibinfo {pages} {1287} (\bibinfo {year} {2015})}\BibitemShut {NoStop}%
\bibitem [{\citenamefont {Schmeltz}\ \emph {et~al.}(2020)\citenamefont
  {Schmeltz}, \citenamefont {Teulon}, \citenamefont {Pinsard}, \citenamefont
  {Hansen}, \citenamefont {Alnawaiseh}, \citenamefont {Ghoubay}, \citenamefont
  {Borderie}, \citenamefont {Mosser}, \citenamefont
  {Aim{\ifmmode\acute{e}\else\'{e}\fi}}, \citenamefont
  {Aim{\ifmmode\acute{e}\else\'{e}\fi}}, \citenamefont
  {L{\ifmmode\acute{e}\else\'{e}\fi}gar{\ifmmode\acute{e}\else\'{e}\fi}},
  \citenamefont {Latour}, \citenamefont {Latour},\ and\ \citenamefont
  {Schanne-Klein}}]{Schmeltz2020-Circulardichroismse}%
  \BibitemOpen
  \bibfield  {author} {\bibinfo {author} {\bibfnamefont {M.}~\bibnamefont
  {Schmeltz}}, \bibinfo {author} {\bibfnamefont {C.}~\bibnamefont {Teulon}},
  \bibinfo {author} {\bibfnamefont {M.}~\bibnamefont {Pinsard}}, \bibinfo
  {author} {\bibfnamefont {U.}~\bibnamefont {Hansen}}, \bibinfo {author}
  {\bibfnamefont {M.}~\bibnamefont {Alnawaiseh}}, \bibinfo {author}
  {\bibfnamefont {D.}~\bibnamefont {Ghoubay}}, \bibinfo {author} {\bibfnamefont
  {V.}~\bibnamefont {Borderie}}, \bibinfo {author} {\bibfnamefont
  {G.}~\bibnamefont {Mosser}}, \bibinfo {author} {\bibfnamefont
  {C.}~\bibnamefont {Aim{\ifmmode\acute{e}\else\'{e}\fi}}}, \bibinfo {author}
  {\bibfnamefont {C.}~\bibnamefont {Aim{\ifmmode\acute{e}\else\'{e}\fi}}},
  \bibinfo {author} {\bibfnamefont {F.}~\bibnamefont
  {L{\ifmmode\acute{e}\else\'{e}\fi}gar{\ifmmode\acute{e}\else\'{e}\fi}}},
  \bibinfo {author} {\bibfnamefont {G.}~\bibnamefont {Latour}}, \bibinfo
  {author} {\bibfnamefont {G.}~\bibnamefont {Latour}},\ and\ \bibinfo {author}
  {\bibfnamefont {M.-C.}\ \bibnamefont {Schanne-Klein}},\ }\bibfield  {title}
  {\bibinfo {title} {{Circular dichroism second-harmonic generation microscopy
  probes the polarity distribution of collagen fibrils}},\ }\href
  {https://doi.org/10.1364/OPTICA.399246} {\bibfield  {journal} {\bibinfo
  {journal} {Optica}\ }\textbf {\bibinfo {volume} {7}},\ \bibinfo {pages}
  {1469} (\bibinfo {year} {2020})}\BibitemShut {NoStop}%
\bibitem [{\citenamefont {Valev}\ \emph {et~al.}(2009)\citenamefont {Valev},
  \citenamefont {Smisdom}, \citenamefont {Silhanek}, \citenamefont {De~Clercq},
  \citenamefont {Gillijns}, \citenamefont {Ameloot}, \citenamefont
  {Moshchalkov},\ and\ \citenamefont {Verbiest}}]{Valev_2009}%
  \BibitemOpen
  \bibfield  {author} {\bibinfo {author} {\bibfnamefont {V.}~\bibnamefont
  {Valev}}, \bibinfo {author} {\bibfnamefont {N.}~\bibnamefont {Smisdom}},
  \bibinfo {author} {\bibfnamefont {A.}~\bibnamefont {Silhanek}}, \bibinfo
  {author} {\bibfnamefont {B.}~\bibnamefont {De~Clercq}}, \bibinfo {author}
  {\bibfnamefont {W.}~\bibnamefont {Gillijns}}, \bibinfo {author}
  {\bibfnamefont {M.}~\bibnamefont {Ameloot}}, \bibinfo {author} {\bibfnamefont
  {V.}~\bibnamefont {Moshchalkov}},\ and\ \bibinfo {author} {\bibfnamefont
  {T.}~\bibnamefont {Verbiest}},\ }\bibfield  {title} {\bibinfo {title}
  {Plasmonic ratchet wheels: Switching circular dichroism by arranging chiral
  nanostructures},\ }\href@noop {} {\bibfield  {journal} {\bibinfo  {journal}
  {Nano Letters}\ }\textbf {\bibinfo {volume} {9}},\ \bibinfo {pages} {3945}
  (\bibinfo {year} {2009})}\BibitemShut {NoStop}%
\bibitem [{\citenamefont {Valev}\ \emph {et~al.}(2010)\citenamefont {Valev},
  \citenamefont {Silhanek}, \citenamefont {Verellen}, \citenamefont {Gillijns},
  \citenamefont {Van~Dorpe}, \citenamefont {Aktsipetrov}, \citenamefont
  {Vandenbosch}, \citenamefont {Moshchalkov},\ and\ \citenamefont
  {Verbiest}}]{PhysRevLett.104.127401}%
  \BibitemOpen
  \bibfield  {author} {\bibinfo {author} {\bibfnamefont {V.~K.}\ \bibnamefont
  {Valev}}, \bibinfo {author} {\bibfnamefont {A.~V.}\ \bibnamefont {Silhanek}},
  \bibinfo {author} {\bibfnamefont {N.}~\bibnamefont {Verellen}}, \bibinfo
  {author} {\bibfnamefont {W.}~\bibnamefont {Gillijns}}, \bibinfo {author}
  {\bibfnamefont {P.}~\bibnamefont {Van~Dorpe}}, \bibinfo {author}
  {\bibfnamefont {O.~A.}\ \bibnamefont {Aktsipetrov}}, \bibinfo {author}
  {\bibfnamefont {G.~A.~E.}\ \bibnamefont {Vandenbosch}}, \bibinfo {author}
  {\bibfnamefont {V.~V.}\ \bibnamefont {Moshchalkov}},\ and\ \bibinfo {author}
  {\bibfnamefont {T.}~\bibnamefont {Verbiest}},\ }\bibfield  {title} {\bibinfo
  {title} {Asymmetric optical second-harmonic generation from chiral $g$-shaped
  gold nanostructures},\ }\href
  {https://doi.org/10.1103/PhysRevLett.104.127401} {\bibfield  {journal}
  {\bibinfo  {journal} {Phys. Rev. Lett.}\ }\textbf {\bibinfo {volume} {104}},\
  \bibinfo {pages} {127401} (\bibinfo {year} {2010})}\BibitemShut {NoStop}%
\bibitem [{\citenamefont {Wang}\ and\ \citenamefont
  {Harutyunyan}(2019)}]{hayk}%
  \BibitemOpen
  \bibfield  {author} {\bibinfo {author} {\bibfnamefont {F.}~\bibnamefont
  {Wang}}\ and\ \bibinfo {author} {\bibfnamefont {H.}~\bibnamefont
  {Harutyunyan}},\ }\bibfield  {title} {\bibinfo {title} {Observation of a
  giant nonlinear chiro-optical response in planar plasmon-photonic
  metasurfaces},\ }\href {https://doi.org/10.1021/acs.nanolett.0c04902}
  {\bibfield  {journal} {\bibinfo  {journal} {Advanced Optical Materials}\
  }\textbf {\bibinfo {volume} {7}},\ \bibinfo {pages} {1900744} (\bibinfo
  {year} {2019})}\BibitemShut {NoStop}%
\bibitem [{\citenamefont {Tymchenko}\ \emph {et~al.}(2015)\citenamefont
  {Tymchenko}, \citenamefont {Gomez-Diaz}, \citenamefont {Lee}, \citenamefont
  {Nookala}, \citenamefont {Belkin},\ and\ \citenamefont
  {Al{\ifmmode\grave{u}\else\`{u}\fi}}}]{Tymchenko2015-GradientNonlinearPa}%
  \BibitemOpen
  \bibfield  {author} {\bibinfo {author} {\bibfnamefont {M.}~\bibnamefont
  {Tymchenko}}, \bibinfo {author} {\bibfnamefont {J.~S.}\ \bibnamefont
  {Gomez-Diaz}}, \bibinfo {author} {\bibfnamefont {J.}~\bibnamefont {Lee}},
  \bibinfo {author} {\bibfnamefont {N.}~\bibnamefont {Nookala}}, \bibinfo
  {author} {\bibfnamefont {M.~A.}\ \bibnamefont {Belkin}},\ and\ \bibinfo
  {author} {\bibfnamefont {A.}~\bibnamefont
  {Al{\ifmmode\grave{u}\else\`{u}\fi}}},\ }\bibfield  {title} {\bibinfo {title}
  {{Gradient Nonlinear Pancharatnam-Berry Metasurfaces}},\ }\href
  {https://doi.org/10.1103/PhysRevLett.115.207403} {\bibfield  {journal}
  {\bibinfo  {journal} {Phys. Rev. Lett.}\ }\textbf {\bibinfo {volume} {115}},\
  \bibinfo {pages} {207403} (\bibinfo {year} {2015})}\BibitemShut {NoStop}%
\bibitem [{\citenamefont {Koshelev}\ and\ \citenamefont
  {Kivshar}(2021)}]{Koshelev_2020}%
  \BibitemOpen
  \bibfield  {author} {\bibinfo {author} {\bibfnamefont {K.}~\bibnamefont
  {Koshelev}}\ and\ \bibinfo {author} {\bibfnamefont {Y.}~\bibnamefont
  {Kivshar}},\ }\bibfield  {title} {\bibinfo {title} {Dielectric resonant
  metaphotonics},\ }\href@noop {} {\bibfield  {journal} {\bibinfo  {journal}
  {ACS Photonics}\ }\textbf {\bibinfo {volume} {8}},\ \bibinfo {pages} {102}
  (\bibinfo {year} {2021})}\BibitemShut {NoStop}%
\bibitem [{\citenamefont {Smirnova}\ and\ \citenamefont
  {Kivshar}(2016)}]{Smirnova_Kivshar_2016}%
  \BibitemOpen
  \bibfield  {author} {\bibinfo {author} {\bibfnamefont {D.}~\bibnamefont
  {Smirnova}}\ and\ \bibinfo {author} {\bibfnamefont {Y.~S.}\ \bibnamefont
  {Kivshar}},\ }\bibfield  {title} {\bibinfo {title} {Multipolar nonlinear
  nanophotonics},\ }\href {https://doi.org/10.1364/OPTICA.3.001241} {\bibfield
  {journal} {\bibinfo  {journal} {Optica}\ }\textbf {\bibinfo {volume} {3}},\
  \bibinfo {pages} {1241} (\bibinfo {year} {2016})}\BibitemShut {NoStop}%
\bibitem [{\citenamefont {Frizyuk}\ \emph
  {et~al.}(2019{\natexlab{a}})\citenamefont {Frizyuk}, \citenamefont
  {Volkovskaya}, \citenamefont {Smirnova}, \citenamefont {Poddubny},\ and\
  \citenamefont {Petrov}}]{Frizyuk_Volkovskaya_Smirnova_Poddubny_Petrov_2019}%
  \BibitemOpen
  \bibfield  {author} {\bibinfo {author} {\bibfnamefont {K.}~\bibnamefont
  {Frizyuk}}, \bibinfo {author} {\bibfnamefont {I.}~\bibnamefont
  {Volkovskaya}}, \bibinfo {author} {\bibfnamefont {D.}~\bibnamefont
  {Smirnova}}, \bibinfo {author} {\bibfnamefont {A.}~\bibnamefont {Poddubny}},\
  and\ \bibinfo {author} {\bibfnamefont {M.}~\bibnamefont {Petrov}},\
  }\bibfield  {title} {\bibinfo {title} {Second-harmonic generation in
  mie-resonant dielectric nanoparticles made of noncentrosymmetric materials},\
  }\href {https://doi.org/10.1103/PhysRevB.99.075425} {\bibfield  {journal}
  {\bibinfo  {journal} {Physical Review B}\ }\textbf {\bibinfo {volume} {99}},\
  \bibinfo {pages} {075425} (\bibinfo {year} {2019}{\natexlab{a}})}\BibitemShut
  {NoStop}%
\bibitem [{\citenamefont {Gigli}\ \emph {et~al.}(2019)\citenamefont {Gigli},
  \citenamefont {Marino}, \citenamefont {Borne}, \citenamefont {Lalanne},\ and\
  \citenamefont {Leo}}]{Gigli2019-All-DielectricNanore}%
  \BibitemOpen
  \bibfield  {author} {\bibinfo {author} {\bibfnamefont {C.}~\bibnamefont
  {Gigli}}, \bibinfo {author} {\bibfnamefont {G.}~\bibnamefont {Marino}},
  \bibinfo {author} {\bibfnamefont {A.}~\bibnamefont {Borne}}, \bibinfo
  {author} {\bibfnamefont {P.}~\bibnamefont {Lalanne}},\ and\ \bibinfo {author}
  {\bibfnamefont {G.}~\bibnamefont {Leo}},\ }\bibfield  {title} {\bibinfo
  {title} {{All-Dielectric Nanoresonators for {$\chi$}(2) Nonlinear Optics}},\
  }\bibfield  {journal} {\bibinfo  {journal} {Front. Phys.}\ }\textbf {\bibinfo
  {volume} {7}},\ \href {https://doi.org/10.3389/fphy.2019.00221}
  {10.3389/fphy.2019.00221} (\bibinfo {year} {2019})\BibitemShut {NoStop}%
\bibitem [{\citenamefont {Renaut}\ \emph {et~al.}(2019)\citenamefont {Renaut},
  \citenamefont {Lang}, \citenamefont {Frizyuk}, \citenamefont {Timofeeva},
  \citenamefont {Komissarenko}, \citenamefont {Mukhin}, \citenamefont
  {Smirnova}, \citenamefont {Timpu}, \citenamefont {Petrov}, \citenamefont
  {Kivshar},\ and\ \citenamefont {Grange}}]{Renaut2019-ReshapingtheSecond}%
  \BibitemOpen
  \bibfield  {author} {\bibinfo {author} {\bibfnamefont {C.}~\bibnamefont
  {Renaut}}, \bibinfo {author} {\bibfnamefont {L.}~\bibnamefont {Lang}},
  \bibinfo {author} {\bibfnamefont {K.}~\bibnamefont {Frizyuk}}, \bibinfo
  {author} {\bibfnamefont {M.}~\bibnamefont {Timofeeva}}, \bibinfo {author}
  {\bibfnamefont {F.~E.}\ \bibnamefont {Komissarenko}}, \bibinfo {author}
  {\bibfnamefont {I.~S.}\ \bibnamefont {Mukhin}}, \bibinfo {author}
  {\bibfnamefont {D.}~\bibnamefont {Smirnova}}, \bibinfo {author}
  {\bibfnamefont {F.}~\bibnamefont {Timpu}}, \bibinfo {author} {\bibfnamefont
  {M.}~\bibnamefont {Petrov}}, \bibinfo {author} {\bibfnamefont
  {Y.}~\bibnamefont {Kivshar}},\ and\ \bibinfo {author} {\bibfnamefont
  {R.}~\bibnamefont {Grange}},\ }\bibfield  {title} {\bibinfo {title}
  {{Reshaping the Second-Order Polar Response of Hybrid
  Metal{\textendash}Dielectric Nanodimers}},\ }\href
  {https://doi.org/10.1021/acs.nanolett.8b04089} {\bibfield  {journal}
  {\bibinfo  {journal} {Nano Lett.}\ }\textbf {\bibinfo {volume} {19}},\
  \bibinfo {pages} {877} (\bibinfo {year} {2019})}\BibitemShut {NoStop}%
\bibitem [{\citenamefont {Kang}\ \emph {et~al.}(2020)\citenamefont {Kang},
  \citenamefont {Wang}, \citenamefont {Guo}, \citenamefont {Ni}, \citenamefont
  {Liu},\ and\ \citenamefont {Werner}}]{Werner2020}%
  \BibitemOpen
  \bibfield  {author} {\bibinfo {author} {\bibfnamefont {L.}~\bibnamefont
  {Kang}}, \bibinfo {author} {\bibfnamefont {C.}~\bibnamefont {Wang}}, \bibinfo
  {author} {\bibfnamefont {X.}~\bibnamefont {Guo}}, \bibinfo {author}
  {\bibfnamefont {X.}~\bibnamefont {Ni}}, \bibinfo {author} {\bibfnamefont
  {Z.}~\bibnamefont {Liu}},\ and\ \bibinfo {author} {\bibfnamefont
  {D.}~\bibnamefont {Werner}},\ }\bibfield  {title} {\bibinfo {title}
  {Nonlinear chiral meta-mirrors: Enabling technology for ultrafast switching
  of light polarisation},\ }\href@noop {} {\bibfield  {journal} {\bibinfo
  {journal} {Nano Letters}\ }\textbf {\bibinfo {volume} {20}},\ \bibinfo
  {pages} {2047} (\bibinfo {year} {2020})}\BibitemShut {NoStop}%
\bibitem [{\citenamefont {Sanatinia}\ \emph {et~al.}(2014)\citenamefont
  {Sanatinia}, \citenamefont {Anand},\ and\ \citenamefont
  {Swillo}}]{Sanatinia_Anand_Swillo_2014}%
  \BibitemOpen
  \bibfield  {author} {\bibinfo {author} {\bibfnamefont {R.}~\bibnamefont
  {Sanatinia}}, \bibinfo {author} {\bibfnamefont {S.}~\bibnamefont {Anand}},\
  and\ \bibinfo {author} {\bibfnamefont {M.}~\bibnamefont {Swillo}},\
  }\bibfield  {title} {\bibinfo {title} {Modal engineering of second-harmonic
  generation in single gap nanopillars},\ }\href
  {https://doi.org/10.1021/nl502521y} {\bibfield  {journal} {\bibinfo
  {journal} {Nano Letters}\ }\textbf {\bibinfo {volume} {14}},\ \bibinfo
  {pages} {5376} (\bibinfo {year} {2014})}\BibitemShut {NoStop}%
\bibitem [{\citenamefont {Carletti}\ \emph {et~al.}(2016)\citenamefont
  {Carletti}, \citenamefont {Locatelli}, \citenamefont {Neshev},\ and\
  \citenamefont {De~Angelis}}]{Carletti2016-ShapingtheRadiation}%
  \BibitemOpen
  \bibfield  {author} {\bibinfo {author} {\bibfnamefont {L.}~\bibnamefont
  {Carletti}}, \bibinfo {author} {\bibfnamefont {A.}~\bibnamefont {Locatelli}},
  \bibinfo {author} {\bibfnamefont {D.}~\bibnamefont {Neshev}},\ and\ \bibinfo
  {author} {\bibfnamefont {C.}~\bibnamefont {De~Angelis}},\ }\bibfield  {title}
  {\bibinfo {title} {{Shaping the Radiation Pattern of Second-Harmonic
  Generation from AlGaAs Dielectric Nanoantennas}},\ }\href
  {https://doi.org/10.1021/acsphotonics.6b00050} {\bibfield  {journal}
  {\bibinfo  {journal} {ACS Photonics}\ }\textbf {\bibinfo {volume} {3}},\
  \bibinfo {pages} {1500} (\bibinfo {year} {2016})}\BibitemShut {NoStop}%
\bibitem [{\citenamefont {Kruk}\ \emph {et~al.}(2017)\citenamefont {Kruk},
  \citenamefont {Camacho-Morales}, \citenamefont {Xu}, \citenamefont {Rahmani},
  \citenamefont {Smirnova}, \citenamefont {Wang}, \citenamefont {Tan},
  \citenamefont {Jagadish}, \citenamefont {Neshev},\ and\ \citenamefont
  {Kivshar}}]{Kruk_Camacho-Morales_Xu_Rahmani_Smirnova_Wang_Tan_Jagadish_Neshev_Kivshar_2017}%
  \BibitemOpen
  \bibfield  {author} {\bibinfo {author} {\bibfnamefont {S.~S.}\ \bibnamefont
  {Kruk}}, \bibinfo {author} {\bibfnamefont {R.}~\bibnamefont
  {Camacho-Morales}}, \bibinfo {author} {\bibfnamefont {L.}~\bibnamefont {Xu}},
  \bibinfo {author} {\bibfnamefont {M.}~\bibnamefont {Rahmani}}, \bibinfo
  {author} {\bibfnamefont {D.~A.}\ \bibnamefont {Smirnova}}, \bibinfo {author}
  {\bibfnamefont {L.}~\bibnamefont {Wang}}, \bibinfo {author} {\bibfnamefont
  {H.~H.}\ \bibnamefont {Tan}}, \bibinfo {author} {\bibfnamefont
  {C.}~\bibnamefont {Jagadish}}, \bibinfo {author} {\bibfnamefont {D.~N.}\
  \bibnamefont {Neshev}},\ and\ \bibinfo {author} {\bibfnamefont {Y.~S.}\
  \bibnamefont {Kivshar}},\ }\bibfield  {title} {\bibinfo {title} {Nonlinear
  optical magnetism revealed by second-harmonic generation in nanoantennas},\
  }\href {https://doi.org/10.1021/acs.nanolett.7b01488} {\bibfield  {journal}
  {\bibinfo  {journal} {Nano Letters}\ }\textbf {\bibinfo {volume} {17}},\
  \bibinfo {pages} {3914} (\bibinfo {year} {2017})}\BibitemShut {NoStop}%
\bibitem [{\citenamefont {Timofeeva}\ \emph {et~al.}(2018)\citenamefont
  {Timofeeva}, \citenamefont {Lang}, \citenamefont {Timpu}, \citenamefont
  {Renaut}, \citenamefont {Bouravleuv}, \citenamefont {Shtrom}, \citenamefont
  {Cirlin},\ and\ \citenamefont
  {Grange}}]{Timofeeva_Lang_Timpu_Renaut_Bouravleuv_Shtrom_Cirlin_Grange_2018}%
  \BibitemOpen
  \bibfield  {author} {\bibinfo {author} {\bibfnamefont {M.}~\bibnamefont
  {Timofeeva}}, \bibinfo {author} {\bibfnamefont {L.}~\bibnamefont {Lang}},
  \bibinfo {author} {\bibfnamefont {F.}~\bibnamefont {Timpu}}, \bibinfo
  {author} {\bibfnamefont {C.}~\bibnamefont {Renaut}}, \bibinfo {author}
  {\bibfnamefont {A.}~\bibnamefont {Bouravleuv}}, \bibinfo {author}
  {\bibfnamefont {I.}~\bibnamefont {Shtrom}}, \bibinfo {author} {\bibfnamefont
  {G.}~\bibnamefont {Cirlin}},\ and\ \bibinfo {author} {\bibfnamefont
  {R.}~\bibnamefont {Grange}},\ }\bibfield  {title} {\bibinfo {title} {Anapoles
  in free-standing iii-v nanodisks enhancing second-harmonic generation},\
  }\href {https://doi.org/10.1021/acs.nanolett.8b00830} {\bibfield  {journal}
  {\bibinfo  {journal} {Nano Letters}\ }\textbf {\bibinfo {volume} {18}},\
  \bibinfo {pages} {3695} (\bibinfo {year} {2018})}\BibitemShut {NoStop}%
\bibitem [{\citenamefont {Sautter}\ \emph {et~al.}(2019)\citenamefont
  {Sautter}, \citenamefont {Xu}, \citenamefont {Miroshnichenko}, \citenamefont
  {Lysevych}, \citenamefont {Volkovskaya}, \citenamefont {Smirnova},
  \citenamefont {Camacho-Morales}, \citenamefont {Zangeneh~Kamali},
  \citenamefont {Karouta}, \citenamefont {Vora}, \citenamefont {Tan},
  \citenamefont {Kauranen}, \citenamefont {Staude}, \citenamefont {Jagadish},
  \citenamefont {Neshev},\ and\ \citenamefont
  {Rahmani}}]{Sautter2019-TailoringSecond-Harm}%
  \BibitemOpen
  \bibfield  {author} {\bibinfo {author} {\bibfnamefont {J.}~\bibnamefont
  {Sautter}}, \bibinfo {author} {\bibfnamefont {L.}~\bibnamefont {Xu}},
  \bibinfo {author} {\bibfnamefont {A.~E.}\ \bibnamefont {Miroshnichenko}},
  \bibinfo {author} {\bibfnamefont {M.}~\bibnamefont {Lysevych}}, \bibinfo
  {author} {\bibfnamefont {I.}~\bibnamefont {Volkovskaya}}, \bibinfo {author}
  {\bibfnamefont {D.~A.}\ \bibnamefont {Smirnova}}, \bibinfo {author}
  {\bibfnamefont {R.}~\bibnamefont {Camacho-Morales}}, \bibinfo {author}
  {\bibfnamefont {K.}~\bibnamefont {Zangeneh~Kamali}}, \bibinfo {author}
  {\bibfnamefont {F.}~\bibnamefont {Karouta}}, \bibinfo {author} {\bibfnamefont
  {K.}~\bibnamefont {Vora}}, \bibinfo {author} {\bibfnamefont {H.~H.}\
  \bibnamefont {Tan}}, \bibinfo {author} {\bibfnamefont {M.}~\bibnamefont
  {Kauranen}}, \bibinfo {author} {\bibfnamefont {I.}~\bibnamefont {Staude}},
  \bibinfo {author} {\bibfnamefont {C.}~\bibnamefont {Jagadish}}, \bibinfo
  {author} {\bibfnamefont {D.~N.}\ \bibnamefont {Neshev}},\ and\ \bibinfo
  {author} {\bibfnamefont {M.}~\bibnamefont {Rahmani}},\ }\bibfield  {title}
  {\bibinfo {title} {{Tailoring Second-Harmonic Emission from (111)-GaAs
  Nanoantennas}},\ }\href {https://doi.org/10.1021/acs.nanolett.9b01112}
  {\bibfield  {journal} {\bibinfo  {journal} {Nano Lett.}\ }\textbf {\bibinfo
  {volume} {19}},\ \bibinfo {pages} {3905} (\bibinfo {year}
  {2019})}\BibitemShut {NoStop}%
\bibitem [{\citenamefont {Koshelev}\ \emph {et~al.}(2020)\citenamefont
  {Koshelev}, \citenamefont {Kruk}, \citenamefont {Melik-Gaykazyan},
  \citenamefont {Choi}, \citenamefont {Bogdanov}, \citenamefont {Park},\ and\
  \citenamefont
  {Kivshar}}]{Koshelev_Kruk_Melik-Gaykazyan_Choi_Bogdanov_Park_Kivshar_2020}%
  \BibitemOpen
  \bibfield  {author} {\bibinfo {author} {\bibfnamefont {K.}~\bibnamefont
  {Koshelev}}, \bibinfo {author} {\bibfnamefont {S.}~\bibnamefont {Kruk}},
  \bibinfo {author} {\bibfnamefont {E.}~\bibnamefont {Melik-Gaykazyan}},
  \bibinfo {author} {\bibfnamefont {J.-H.}\ \bibnamefont {Choi}}, \bibinfo
  {author} {\bibfnamefont {A.}~\bibnamefont {Bogdanov}}, \bibinfo {author}
  {\bibfnamefont {H.-G.}\ \bibnamefont {Park}},\ and\ \bibinfo {author}
  {\bibfnamefont {Y.}~\bibnamefont {Kivshar}},\ }\bibfield  {title} {\bibinfo
  {title} {Subwavelength dielectric resonators for nonlinear nanophotonics},\
  }\href {https://doi.org/10.1126/science.aaz3985} {\bibfield  {journal}
  {\bibinfo  {journal} {Science}\ }\textbf {\bibinfo {volume} {367}},\ \bibinfo
  {pages} {288} (\bibinfo {year} {2020})}\BibitemShut {NoStop}%
\bibitem [{\citenamefont {Boyd}(2003)}]{Boyd2003}%
  \BibitemOpen
  \bibfield  {author} {\bibinfo {author} {\bibfnamefont {R.~W.}\ \bibnamefont
  {Boyd}},\ }\href@noop {} {\emph {\bibinfo {title} {Nonlinear optics}}}\
  (\bibinfo  {publisher} {Academic press},\ \bibinfo {year} {2003})\BibitemShut
  {NoStop}%
\bibitem [{\citenamefont {Novotny}\ and\ \citenamefont
  {Hecht}(2006)}]{Novotny}%
  \BibitemOpen
  \bibfield  {author} {\bibinfo {author} {\bibfnamefont {L.}~\bibnamefont
  {Novotny}}\ and\ \bibinfo {author} {\bibfnamefont {B.}~\bibnamefont
  {Hecht}},\ }\href@noop {} {\emph {\bibinfo {title} {Principles of
  Nano-Optics}}}\ (\bibinfo  {publisher} {Cambridge University Press},\
  \bibinfo {address} {USA},\ \bibinfo {year} {2006})\BibitemShut {NoStop}%
\bibitem [{\citenamefont {Doost}\ \emph {et~al.}(2014)\citenamefont {Doost},
  \citenamefont {Langbein},\ and\ \citenamefont
  {Muljarov}}]{PhysRevA.90.013834}%
  \BibitemOpen
  \bibfield  {author} {\bibinfo {author} {\bibfnamefont {M.~B.}\ \bibnamefont
  {Doost}}, \bibinfo {author} {\bibfnamefont {W.}~\bibnamefont {Langbein}},\
  and\ \bibinfo {author} {\bibfnamefont {E.~A.}\ \bibnamefont {Muljarov}},\
  }\bibfield  {title} {\bibinfo {title} {Resonant-state expansion applied to
  three-dimensional open optical systems},\ }\href
  {https://doi.org/10.1103/PhysRevA.90.013834} {\bibfield  {journal} {\bibinfo
  {journal} {Phys. Rev. A}\ }\textbf {\bibinfo {volume} {90}},\ \bibinfo
  {pages} {013834} (\bibinfo {year} {2014})}\BibitemShut {NoStop}%
\bibitem [{\citenamefont {Gigli}\ \emph {et~al.}(2020)\citenamefont {Gigli},
  \citenamefont {Wu}, \citenamefont {Marino}, \citenamefont {Borne},
  \citenamefont {Leo},\ and\ \citenamefont
  {Lalanne}}]{Gigli2020-Quasinormal-ModeNon}%
  \BibitemOpen
  \bibfield  {author} {\bibinfo {author} {\bibfnamefont {C.}~\bibnamefont
  {Gigli}}, \bibinfo {author} {\bibfnamefont {T.}~\bibnamefont {Wu}}, \bibinfo
  {author} {\bibfnamefont {G.}~\bibnamefont {Marino}}, \bibinfo {author}
  {\bibfnamefont {A.}~\bibnamefont {Borne}}, \bibinfo {author} {\bibfnamefont
  {G.}~\bibnamefont {Leo}},\ and\ \bibinfo {author} {\bibfnamefont
  {P.}~\bibnamefont {Lalanne}},\ }\bibfield  {title} {\bibinfo {title}
  {{Quasinormal-Mode Non-Hermitian Modeling and Design in Nonlinear
  Nano-Optics}},\ }\href {https://doi.org/10.1021/acsphotonics.0c00014}
  {\bibfield  {journal} {\bibinfo  {journal} {ACS Photonics}\ }\textbf
  {\bibinfo {volume} {7}},\ \bibinfo {pages} {1197} (\bibinfo {year}
  {2020})}\BibitemShut {NoStop}%
\bibitem [{\citenamefont {Gladyshev}\ \emph {et~al.}(2020)\citenamefont
  {Gladyshev}, \citenamefont {Frizyuk},\ and\ \citenamefont
  {Bogdanov}}]{Gladyshev_Frizyuk_Bogdanov_2020}%
  \BibitemOpen
  \bibfield  {author} {\bibinfo {author} {\bibfnamefont {S.}~\bibnamefont
  {Gladyshev}}, \bibinfo {author} {\bibfnamefont {K.}~\bibnamefont {Frizyuk}},\
  and\ \bibinfo {author} {\bibfnamefont {A.}~\bibnamefont {Bogdanov}},\
  }\bibfield  {title} {\bibinfo {title} {Symmetry analysis and multipole
  classification of eigenmodes in electromagnetic resonators for engineering
  their optical properties},\ }\href
  {https://doi.org/10.1103/PhysRevB.102.075103} {\bibfield  {journal} {\bibinfo
   {journal} {Physical Review B}\ }\textbf {\bibinfo {volume} {102}},\ \bibinfo
  {pages} {075103} (\bibinfo {year} {2020})}\BibitemShut {NoStop}%
\bibitem [{\citenamefont {Brandl}\ \emph {et~al.}(2006)\citenamefont {Brandl},
  \citenamefont {Mirin},\ and\ \citenamefont
  {Nordlander}}]{Brandl_Mirin_Nordlander_2006}%
  \BibitemOpen
  \bibfield  {author} {\bibinfo {author} {\bibfnamefont {D.~W.}\ \bibnamefont
  {Brandl}}, \bibinfo {author} {\bibfnamefont {N.~A.}\ \bibnamefont {Mirin}},\
  and\ \bibinfo {author} {\bibfnamefont {P.}~\bibnamefont {Nordlander}},\
  }\bibfield  {title} {\bibinfo {title} {Plasmon modes of nanosphere trimers
  and quadrumers},\ }\href {https://doi.org/10.1021/jp0613485} {\bibfield
  {journal} {\bibinfo  {journal} {The Journal of Physical Chemistry B}\
  }\textbf {\bibinfo {volume} {110}},\ \bibinfo {pages} {12302} (\bibinfo
  {year} {2006})}\BibitemShut {NoStop}%
\bibitem [{\citenamefont {Zheng}\ \emph {et~al.}(2015)\citenamefont {Zheng},
  \citenamefont {Verellen}, \citenamefont {Vercruysse}, \citenamefont
  {Volskiy}, \citenamefont {Van~Dorpe}, \citenamefont {Vandenbosch},\ and\
  \citenamefont {Moshchalkov}}]{Zheng2015-OntheUseofGroupT}%
  \BibitemOpen
  \bibfield  {author} {\bibinfo {author} {\bibfnamefont {X.}~\bibnamefont
  {Zheng}}, \bibinfo {author} {\bibfnamefont {N.}~\bibnamefont {Verellen}},
  \bibinfo {author} {\bibfnamefont {D.}~\bibnamefont {Vercruysse}}, \bibinfo
  {author} {\bibfnamefont {V.}~\bibnamefont {Volskiy}}, \bibinfo {author}
  {\bibfnamefont {P.}~\bibnamefont {Van~Dorpe}}, \bibinfo {author}
  {\bibfnamefont {G.~A.~E.}\ \bibnamefont {Vandenbosch}},\ and\ \bibinfo
  {author} {\bibfnamefont {V.}~\bibnamefont {Moshchalkov}},\ }\bibfield
  {title} {\bibinfo {title} {{On the Use of Group Theory in Understanding the
  Optical Response of a Nanoantenna}},\ }\href
  {https://doi.org/10.1109/TAP.2015.2400471} {\bibfield  {journal} {\bibinfo
  {journal} {IEEE Trans. Antennas Propag.}\ }\textbf {\bibinfo {volume} {63}},\
  \bibinfo {pages} {1589} (\bibinfo {year} {2015})}\BibitemShut {NoStop}%
\bibitem [{\citenamefont {Chikkaraddy}\ \emph {et~al.}(2017)\citenamefont
  {Chikkaraddy}, \citenamefont {Zheng}, \citenamefont {Benz}, \citenamefont
  {Brooks}, \citenamefont {de~Nijs}, \citenamefont {Carnegie}, \citenamefont
  {Kleemann}, \citenamefont {Mertens}, \citenamefont {Bowman}, \citenamefont
  {Vandenbosch}, \citenamefont {Moshchalkov},\ and\ \citenamefont
  {Baumberg}}]{Chikkaraddy2017-HowUltranarrowGapS}%
  \BibitemOpen
  \bibfield  {author} {\bibinfo {author} {\bibfnamefont {R.}~\bibnamefont
  {Chikkaraddy}}, \bibinfo {author} {\bibfnamefont {X.}~\bibnamefont {Zheng}},
  \bibinfo {author} {\bibfnamefont {F.}~\bibnamefont {Benz}}, \bibinfo {author}
  {\bibfnamefont {L.~J.}\ \bibnamefont {Brooks}}, \bibinfo {author}
  {\bibfnamefont {B.}~\bibnamefont {de~Nijs}}, \bibinfo {author} {\bibfnamefont
  {C.}~\bibnamefont {Carnegie}}, \bibinfo {author} {\bibfnamefont {M.-E.}\
  \bibnamefont {Kleemann}}, \bibinfo {author} {\bibfnamefont {J.}~\bibnamefont
  {Mertens}}, \bibinfo {author} {\bibfnamefont {R.~W.}\ \bibnamefont {Bowman}},
  \bibinfo {author} {\bibfnamefont {G.~A.~E.}\ \bibnamefont {Vandenbosch}},
  \bibinfo {author} {\bibfnamefont {V.~V.}\ \bibnamefont {Moshchalkov}},\ and\
  \bibinfo {author} {\bibfnamefont {J.~J.}\ \bibnamefont {Baumberg}},\
  }\bibfield  {title} {\bibinfo {title} {{How Ultranarrow Gap Symmetries
  Control Plasmonic Nanocavity Modes: From Cubes to Spheres in the
  Nanoparticle-on-Mirror}},\ }\href
  {https://doi.org/10.1021/acsphotonics.6b00908} {\bibfield  {journal}
  {\bibinfo  {journal} {ACS Photonics}\ }\textbf {\bibinfo {volume} {4}},\
  \bibinfo {pages} {469} (\bibinfo {year} {2017})}\BibitemShut {NoStop}%
\bibitem [{\citenamefont {Xiong}\ \emph {et~al.}(2020)\citenamefont {Xiong},
  \citenamefont {Xiong}, \citenamefont {Yang}, \citenamefont {Yang},
  \citenamefont {Chen}, \citenamefont {Wang}, \citenamefont {Xu}, \citenamefont
  {Xu}, \citenamefont {Xu}, \citenamefont {Liu},\ and\ \citenamefont
  {et~al.}}]{Xiong_Xiong_Yang_Yang_Chen_Wang_Xu_Xu_Xu_Liu_2020}%
  \BibitemOpen
  \bibfield  {author} {\bibinfo {author} {\bibfnamefont {Z.}~\bibnamefont
  {Xiong}}, \bibinfo {author} {\bibfnamefont {Z.}~\bibnamefont {Xiong}},
  \bibinfo {author} {\bibfnamefont {Q.}~\bibnamefont {Yang}}, \bibinfo {author}
  {\bibfnamefont {Q.}~\bibnamefont {Yang}}, \bibinfo {author} {\bibfnamefont
  {W.}~\bibnamefont {Chen}}, \bibinfo {author} {\bibfnamefont {Z.}~\bibnamefont
  {Wang}}, \bibinfo {author} {\bibfnamefont {J.}~\bibnamefont {Xu}}, \bibinfo
  {author} {\bibfnamefont {J.}~\bibnamefont {Xu}}, \bibinfo {author}
  {\bibfnamefont {J.}~\bibnamefont {Xu}}, \bibinfo {author} {\bibfnamefont
  {W.}~\bibnamefont {Liu}},\ and\ \bibinfo {author} {\bibnamefont {et~al.}},\
  }\bibfield  {title} {\bibinfo {title} {On the constraints of electromagnetic
  multipoles for symmetric scatterers: eigenmode analysis},\ }\href
  {https://doi.org/10.1364/OE.382239} {\bibfield  {journal} {\bibinfo
  {journal} {Optics Express}\ }\textbf {\bibinfo {volume} {28}},\ \bibinfo
  {pages} {3073} (\bibinfo {year} {2020})}\BibitemShut {NoStop}%
\bibitem [{\citenamefont {Horrer}\ \emph {et~al.}(2020)\citenamefont {Horrer},
  \citenamefont {Zhang}, \citenamefont {G{\ifmmode\acute{e}\else\'{e}\fi}rard},
  \citenamefont {B{\ifmmode\acute{e}\else\'{e}\fi}al}, \citenamefont {Kociak},
  \citenamefont {Plain},\ and\ \citenamefont
  {Bachelot}}]{Horrer2020-LocalOpticalChirali}%
  \BibitemOpen
  \bibfield  {author} {\bibinfo {author} {\bibfnamefont {A.}~\bibnamefont
  {Horrer}}, \bibinfo {author} {\bibfnamefont {Y.}~\bibnamefont {Zhang}},
  \bibinfo {author} {\bibfnamefont {D.}~\bibnamefont
  {G{\ifmmode\acute{e}\else\'{e}\fi}rard}}, \bibinfo {author} {\bibfnamefont
  {J.}~\bibnamefont {B{\ifmmode\acute{e}\else\'{e}\fi}al}}, \bibinfo {author}
  {\bibfnamefont {M.}~\bibnamefont {Kociak}}, \bibinfo {author} {\bibfnamefont
  {J.}~\bibnamefont {Plain}},\ and\ \bibinfo {author} {\bibfnamefont
  {R.}~\bibnamefont {Bachelot}},\ }\bibfield  {title} {\bibinfo {title} {{Local
  Optical Chirality Induced by Near-Field Mode Interference in Achiral
  Plasmonic Metamolecules}},\ }\href
  {https://doi.org/10.1021/acs.nanolett.9b04247} {\bibfield  {journal}
  {\bibinfo  {journal} {Nano Lett.}\ }\textbf {\bibinfo {volume} {20}},\
  \bibinfo {pages} {509} (\bibinfo {year} {2020})}\BibitemShut {NoStop}%
\bibitem [{\citenamefont {Zu}\ \emph {et~al.}(2018)\citenamefont {Zu},
  \citenamefont {Han}, \citenamefont {Jiang}, \citenamefont {Lin},
  \citenamefont {Zhu},\ and\ \citenamefont
  {Fang}}]{Zu2018-Deep-SubwavelengthRe}%
  \BibitemOpen
  \bibfield  {author} {\bibinfo {author} {\bibfnamefont {S.}~\bibnamefont
  {Zu}}, \bibinfo {author} {\bibfnamefont {T.}~\bibnamefont {Han}}, \bibinfo
  {author} {\bibfnamefont {M.}~\bibnamefont {Jiang}}, \bibinfo {author}
  {\bibfnamefont {F.}~\bibnamefont {Lin}}, \bibinfo {author} {\bibfnamefont
  {X.}~\bibnamefont {Zhu}},\ and\ \bibinfo {author} {\bibfnamefont
  {Z.}~\bibnamefont {Fang}},\ }\bibfield  {title} {\bibinfo {title}
  {{Deep-Subwavelength Resolving and Manipulating of Hidden Chirality in
  Achiral Nanostructures}},\ }\href {https://doi.org/10.1021/acsnano.8b01380}
  {\bibfield  {journal} {\bibinfo  {journal} {ACS Nano}\ }\textbf {\bibinfo
  {volume} {12}},\ \bibinfo {pages} {3908} (\bibinfo {year}
  {2018})}\BibitemShut {NoStop}%
\bibitem [{\citenamefont {Nordlander}\ \emph {et~al.}(2004)\citenamefont
  {Nordlander}, \citenamefont {Oubre}, \citenamefont {Prodan}, \citenamefont
  {Li},\ and\ \citenamefont {Stockman}}]{Nordlander2004-PlasmonHybridization}%
  \BibitemOpen
  \bibfield  {author} {\bibinfo {author} {\bibfnamefont {P.}~\bibnamefont
  {Nordlander}}, \bibinfo {author} {\bibfnamefont {C.}~\bibnamefont {Oubre}},
  \bibinfo {author} {\bibfnamefont {E.}~\bibnamefont {Prodan}}, \bibinfo
  {author} {\bibfnamefont {K.}~\bibnamefont {Li}},\ and\ \bibinfo {author}
  {\bibfnamefont {M.~I.}\ \bibnamefont {Stockman}},\ }\bibfield  {title}
  {\bibinfo {title} {{Plasmon Hybridization in Nanoparticle Dimers}},\ }\href
  {https://doi.org/10.1021/nl049681c} {\bibfield  {journal} {\bibinfo
  {journal} {Nano Lett.}\ }\textbf {\bibinfo {volume} {4}},\ \bibinfo {pages}
  {899} (\bibinfo {year} {2004})}\BibitemShut {NoStop}%
\bibitem [{\citenamefont {Gao}\ \emph {et~al.}(2017)\citenamefont {Gao},
  \citenamefont {Gao}, \citenamefont {Zhang}, \citenamefont {Xu}, \citenamefont
  {Luo},\ and\ \citenamefont {Zhang}}]{Gao2017-ForwardBackwardSwit}%
  \BibitemOpen
  \bibfield  {author} {\bibinfo {author} {\bibfnamefont {Z.}~\bibnamefont
  {Gao}}, \bibinfo {author} {\bibfnamefont {F.}~\bibnamefont {Gao}}, \bibinfo
  {author} {\bibfnamefont {Y.}~\bibnamefont {Zhang}}, \bibinfo {author}
  {\bibfnamefont {H.}~\bibnamefont {Xu}}, \bibinfo {author} {\bibfnamefont
  {Y.}~\bibnamefont {Luo}},\ and\ \bibinfo {author} {\bibfnamefont
  {B.}~\bibnamefont {Zhang}},\ }\bibfield  {title} {\bibinfo {title}
  {{Forward/Backward Switching of Plasmonic Wave Propagation Using
  Sign-Reversal Coupling}},\ }\href {https://doi.org/10.1002/adma.201700018}
  {\bibfield  {journal} {\bibinfo  {journal} {Adv. Mater.}\ }\textbf {\bibinfo
  {volume} {29}},\ \bibinfo {pages} {1700018} (\bibinfo {year}
  {2017})}\BibitemShut {NoStop}%
\bibitem [{\citenamefont {Deng}\ \emph {et~al.}(2018)\citenamefont {Deng},
  \citenamefont {Parker}, \citenamefont {Yifat}, \citenamefont {Shepherd},\
  and\ \citenamefont {Scherer}}]{Deng2018-DarkPlasmonModesin}%
  \BibitemOpen
  \bibfield  {author} {\bibinfo {author} {\bibfnamefont {T.-S.}\ \bibnamefont
  {Deng}}, \bibinfo {author} {\bibfnamefont {J.}~\bibnamefont {Parker}},
  \bibinfo {author} {\bibfnamefont {Y.}~\bibnamefont {Yifat}}, \bibinfo
  {author} {\bibfnamefont {N.}~\bibnamefont {Shepherd}},\ and\ \bibinfo
  {author} {\bibfnamefont {N.~F.}\ \bibnamefont {Scherer}},\ }\bibfield
  {title} {\bibinfo {title} {{Dark Plasmon Modes in Symmetric Gold Nanoparticle
  Dimers Illuminated by Focused Cylindrical Vector Beams}},\ }\href
  {https://doi.org/10.1021/acs.jpcc.8b10415} {\bibfield  {journal} {\bibinfo
  {journal} {J. Phys. Chem. C}\ }\textbf {\bibinfo {volume} {122}},\ \bibinfo
  {pages} {27662} (\bibinfo {year} {2018})}\BibitemShut {NoStop}%
\bibitem [{\citenamefont {Pascale}\ \emph {et~al.}(2019)\citenamefont
  {Pascale}, \citenamefont {Miano}, \citenamefont {Tricarico},\ and\
  \citenamefont {Forestiere}}]{Pascale2019-Full-waveelectromagn}%
  \BibitemOpen
  \bibfield  {author} {\bibinfo {author} {\bibfnamefont {M.}~\bibnamefont
  {Pascale}}, \bibinfo {author} {\bibfnamefont {G.}~\bibnamefont {Miano}},
  \bibinfo {author} {\bibfnamefont {R.}~\bibnamefont {Tricarico}},\ and\
  \bibinfo {author} {\bibfnamefont {C.}~\bibnamefont {Forestiere}},\ }\bibfield
   {title} {\bibinfo {title} {{Full-wave electromagnetic modes and
  hybridization in nanoparticle dimers}},\ }\href
  {https://doi.org/10.1038/s41598-019-50498-1} {\bibfield  {journal} {\bibinfo
  {journal} {Sci. Rep.}\ }\textbf {\bibinfo {volume} {9}},\ \bibinfo {pages}
  {1} (\bibinfo {year} {2019})}\BibitemShut {NoStop}%
\bibitem [{\citenamefont {Hopkins}\ \emph {et~al.}(2015)\citenamefont
  {Hopkins}, \citenamefont {Filonov}, \citenamefont {Glybovski},\ and\
  \citenamefont {Miroshnichenko}}]{PhysRevB.92.045433}%
  \BibitemOpen
  \bibfield  {author} {\bibinfo {author} {\bibfnamefont {B.}~\bibnamefont
  {Hopkins}}, \bibinfo {author} {\bibfnamefont {D.~S.}\ \bibnamefont
  {Filonov}}, \bibinfo {author} {\bibfnamefont {S.~B.}\ \bibnamefont
  {Glybovski}},\ and\ \bibinfo {author} {\bibfnamefont {A.~E.}\ \bibnamefont
  {Miroshnichenko}},\ }\bibfield  {title} {\bibinfo {title} {Hybridization and
  the origin of fano resonances in symmetric nanoparticle trimers},\ }\href
  {https://doi.org/10.1103/PhysRevB.92.045433} {\bibfield  {journal} {\bibinfo
  {journal} {Phys. Rev. B}\ }\textbf {\bibinfo {volume} {92}},\ \bibinfo
  {pages} {045433} (\bibinfo {year} {2015})}\BibitemShut {NoStop}%
\bibitem [{\citenamefont {Dmitriev}\ \emph {et~al.}(2021)\citenamefont
  {Dmitriev}, \citenamefont {Rybin},\ and\ \citenamefont
  {Rybin}}]{Dmitriev2021-Opticalcouplingofo}%
  \BibitemOpen
  \bibfield  {author} {\bibinfo {author} {\bibfnamefont {A.~A.}\ \bibnamefont
  {Dmitriev}}, \bibinfo {author} {\bibfnamefont {M.~V.}\ \bibnamefont
  {Rybin}},\ and\ \bibinfo {author} {\bibfnamefont {M.~V.}\ \bibnamefont
  {Rybin}},\ }\bibfield  {title} {\bibinfo {title} {{Optical coupling of
  overlapping nanopillars}},\ }\href {https://doi.org/10.1364/OL.415334}
  {\bibfield  {journal} {\bibinfo  {journal} {Opt. Lett.}\ }\textbf {\bibinfo
  {volume} {46}},\ \bibinfo {pages} {1221} (\bibinfo {year}
  {2021})}\BibitemShut {NoStop}%
\bibitem [{\citenamefont {Song}\ \emph {et~al.}(2021)\citenamefont {Song},
  \citenamefont {Raza}, \citenamefont {van~de Groep}, \citenamefont {Kang},
  \citenamefont {Li}, \citenamefont {Kik},\ and\ \citenamefont
  {Brongersma}}]{Song2021-Nanoelectromechanical}%
  \BibitemOpen
  \bibfield  {author} {\bibinfo {author} {\bibfnamefont {J.-H.}\ \bibnamefont
  {Song}}, \bibinfo {author} {\bibfnamefont {S.}~\bibnamefont {Raza}}, \bibinfo
  {author} {\bibfnamefont {J.}~\bibnamefont {van~de Groep}}, \bibinfo {author}
  {\bibfnamefont {J.-H.}\ \bibnamefont {Kang}}, \bibinfo {author}
  {\bibfnamefont {Q.}~\bibnamefont {Li}}, \bibinfo {author} {\bibfnamefont
  {P.~G.}\ \bibnamefont {Kik}},\ and\ \bibinfo {author} {\bibfnamefont {M.~L.}\
  \bibnamefont {Brongersma}},\ }\bibfield  {title} {\bibinfo {title}
  {{Nanoelectromechanical modulation of a strongly-coupled plasmonic dimer}},\
  }\href {https://doi.org/10.1038/s41467-020-20273-2} {\bibfield  {journal}
  {\bibinfo  {journal} {Nat. Commun.}\ }\textbf {\bibinfo {volume} {12}},\
  \bibinfo {pages} {1} (\bibinfo {year} {2021})}\BibitemShut {NoStop}%
\bibitem [{\citenamefont {Frizyuk}(2019)}]{Frizyuk_2019}%
  \BibitemOpen
  \bibfield  {author} {\bibinfo {author} {\bibfnamefont {K.}~\bibnamefont
  {Frizyuk}},\ }\bibfield  {title} {\bibinfo {title} {Second-harmonic
  generation in dielectric nanoparticles with different symmetries},\ }\href
  {https://doi.org/10.1364/JOSAB.36.000F32} {\bibfield  {journal} {\bibinfo
  {journal} {JOSA B}\ }\textbf {\bibinfo {volume} {36}},\ \bibinfo {pages}
  {F32} (\bibinfo {year} {2019})}\BibitemShut {NoStop}%
\bibitem [{\citenamefont {Melik-Gaykazyan}\ \emph {et~al.}(2021)\citenamefont
  {Melik-Gaykazyan}, \citenamefont {Koshelev}, \citenamefont {Choi},
  \citenamefont {Kruk}, \citenamefont {Bogdanov}, \citenamefont {Park},\ and\
  \citenamefont {Kivshar}}]{melik2021}%
  \BibitemOpen
  \bibfield  {author} {\bibinfo {author} {\bibfnamefont {E.}~\bibnamefont
  {Melik-Gaykazyan}}, \bibinfo {author} {\bibfnamefont {K.}~\bibnamefont
  {Koshelev}}, \bibinfo {author} {\bibfnamefont {J.-H.}\ \bibnamefont {Choi}},
  \bibinfo {author} {\bibfnamefont {S.~S.}\ \bibnamefont {Kruk}}, \bibinfo
  {author} {\bibfnamefont {A.}~\bibnamefont {Bogdanov}}, \bibinfo {author}
  {\bibfnamefont {H.-G.}\ \bibnamefont {Park}},\ and\ \bibinfo {author}
  {\bibfnamefont {Y.}~\bibnamefont {Kivshar}},\ }\bibfield  {title} {\bibinfo
  {title} {From fano to quasi-bic resonances in individual dielectric
  nanoantennas},\ }\href {https://doi.org/10.1021/acs.nanolett.0c04660}
  {\bibfield  {journal} {\bibinfo  {journal} {Nano Letters}\ }\textbf {\bibinfo
  {volume} {21}},\ \bibinfo {pages} {1765} (\bibinfo {year}
  {2021})}\BibitemShut {NoStop}%
\bibitem [{\citenamefont {STEIN}(1961)}]{10.2307/43634833}%
  \BibitemOpen
  \bibfield  {author} {\bibinfo {author} {\bibfnamefont {S.}~\bibnamefont
  {STEIN}},\ }\bibfield  {title} {\bibinfo {title} {Addition theorems for
  spherical wave functions},\ }\href {http://www.jstor.org/stable/43634833}
  {\bibfield  {journal} {\bibinfo  {journal} {Quarterly of Applied
  Mathematics}\ }\textbf {\bibinfo {volume} {19}},\ \bibinfo {pages} {15}
  (\bibinfo {year} {1961})}\BibitemShut {NoStop}%
\bibitem [{\citenamefont {{Contributors to Wikimedia
  projects}}(2021)}]{ContributorstoWikimediaprojects2021-Vectorsphe}%
  \BibitemOpen
  \bibfield  {author} {\bibinfo {author} {\bibnamefont {{Contributors to
  Wikimedia projects}}},\ }\href
  {https://en.wikipedia.org/w/index.php?title=Vector_spherical_harmonics&oldid=1008473655}
  {\bibinfo {title} {{Vector spherical harmonics}}} (\bibinfo {year} {2021}),\
  \bibinfo {note} {[Online; accessed 9. Mar. 2021]}\BibitemShut {NoStop}%
\bibitem [{\citenamefont {Zhang}\ and\ \citenamefont
  {Han}(2008)}]{Zhang2008-Additiontheoremfor}%
  \BibitemOpen
  \bibfield  {author} {\bibinfo {author} {\bibfnamefont {H.}~\bibnamefont
  {Zhang}}\ and\ \bibinfo {author} {\bibfnamefont {Y.}~\bibnamefont {Han}},\
  }\bibfield  {title} {\bibinfo {title} {{Addition theorem for the spherical
  vector wave functions and its application to the beam shape coefficients}},\
  }\href {https://doi.org/10.1364/JOSAB.25.000255} {\bibfield  {journal}
  {\bibinfo  {journal} {J. Opt. Soc. Am. B, JOSAB}\ }\textbf {\bibinfo {volume}
  {25}},\ \bibinfo {pages} {255} (\bibinfo {year} {2008})}\BibitemShut
  {NoStop}%
\bibitem [{\citenamefont {Venkatesh}\ and\ \citenamefont
  {Schurig}(2016)}]{Venkatesh_Schurig_2016}%
  \BibitemOpen
  \bibfield  {author} {\bibinfo {author} {\bibfnamefont {S.}~\bibnamefont
  {Venkatesh}}\ and\ \bibinfo {author} {\bibfnamefont {D.}~\bibnamefont
  {Schurig}},\ }\bibfield  {title} {\bibinfo {title} {Computationally fast em
  field propagation through axi-symmetric media using cylindrical harmonic
  decomposition},\ }\href {https://doi.org/10.1364/OE.24.029246} {\bibfield
  {journal} {\bibinfo  {journal} {Optics Express}\ }\textbf {\bibinfo {volume}
  {24}},\ \bibinfo {pages} {29246} (\bibinfo {year} {2016})}\BibitemShut
  {NoStop}%
\bibitem [{\citenamefont {Bohren}\ and\ \citenamefont
  {Huffman}(1983)}]{Bohren}%
  \BibitemOpen
  \bibfield  {author} {\bibinfo {author} {\bibfnamefont {C.~F.}\ \bibnamefont
  {Bohren}}\ and\ \bibinfo {author} {\bibfnamefont {D.~R.}\ \bibnamefont
  {Huffman}},\ }\href {https://doi.org/10.1017/S0263574798270858} {\emph
  {\bibinfo {title} {Absorption and scattering of light by small particles.}}}\
  (\bibinfo  {publisher} {New York Wiley Interscience},\ \bibinfo {year}
  {1983})\ pp.\ \bibinfo {pages} {xiv, 530 p.}\BibitemShut {Stop}%
\bibitem [{\citenamefont {Dresselhaus}\ \emph {et~al.}(2008)\citenamefont
  {Dresselhaus}, \citenamefont {Dresselhaus},\ and\ \citenamefont
  {Jorio}}]{Dresselhaus}%
  \BibitemOpen
  \bibfield  {author} {\bibinfo {author} {\bibfnamefont {M.~S.}\ \bibnamefont
  {Dresselhaus}}, \bibinfo {author} {\bibfnamefont {G.}~\bibnamefont
  {Dresselhaus}},\ and\ \bibinfo {author} {\bibfnamefont {A.}~\bibnamefont
  {Jorio}},\ }\href {https://www.springer.com/gp/book/9783540328971} {\emph
  {\bibinfo {title} {{Group Theory}}}}\ (\bibinfo  {publisher}
  {Springer-Verlag},\ \bibinfo {address} {Berlin, Germany},\ \bibinfo {year}
  {2008})\BibitemShut {NoStop}%
\bibitem [{\citenamefont {Ivchenko}\ \emph {et~al.}(2012)\citenamefont
  {Ivchenko}, \citenamefont {Skrebtsov},\ and\ \citenamefont
  {Pikus}}]{Ivchenko2012-SuperlatticesandOth}%
  \BibitemOpen
  \bibfield  {author} {\bibinfo {author} {\bibfnamefont {E.~L.}\ \bibnamefont
  {Ivchenko}}, \bibinfo {author} {\bibfnamefont {G.~P.}\ \bibnamefont
  {Skrebtsov}},\ and\ \bibinfo {author} {\bibfnamefont {G.}~\bibnamefont
  {Pikus}},\ }\href
  {https://books.google.ru/books?id=u2frCAAAQBAJ&pg=PA341&lpg=PA341&dq=ivchenko+group+theory&source=bl&ots=XwF2PLNNri&sig=ACfU3U3bB_nXjrzt7Z4HAJA08PGUTX3aFg&hl=ru&sa=X&ved=2ahUKEwj1opqx6rrvAhWXHXcKHe56As0Q6AEwFHoECAoQAw#v=onepage&q=ivchenko\%20group\%20theory&f=false}
  {\emph {\bibinfo {title} {{Superlattices and Other Heterostructures}}}}\
  (\bibinfo  {publisher} {Springer},\ \bibinfo {address} {Berlin, Germany},\
  \bibinfo {year} {2012})\BibitemShut {NoStop}%
\bibitem [{\citenamefont {Frizyuk}\ \emph
  {et~al.}(2019{\natexlab{b}})\citenamefont {Frizyuk}, \citenamefont
  {Volkovskaya}, \citenamefont {Smirnova}, \citenamefont {Poddubny},\ and\
  \citenamefont {Petrov}}]{PhysRevB.99.075425}%
  \BibitemOpen
  \bibfield  {author} {\bibinfo {author} {\bibfnamefont {K.}~\bibnamefont
  {Frizyuk}}, \bibinfo {author} {\bibfnamefont {I.}~\bibnamefont
  {Volkovskaya}}, \bibinfo {author} {\bibfnamefont {D.}~\bibnamefont
  {Smirnova}}, \bibinfo {author} {\bibfnamefont {A.}~\bibnamefont {Poddubny}},\
  and\ \bibinfo {author} {\bibfnamefont {M.}~\bibnamefont {Petrov}},\
  }\bibfield  {title} {\bibinfo {title} {Second-harmonic generation in
  mie-resonant dielectric nanoparticles made of noncentrosymmetric materials},\
  }\href {https://doi.org/10.1103/PhysRevB.99.075425} {\bibfield  {journal}
  {\bibinfo  {journal} {Phys. Rev. B}\ }\textbf {\bibinfo {volume} {99}},\
  \bibinfo {pages} {075425} (\bibinfo {year} {2019}{\natexlab{b}})}\BibitemShut
  {NoStop}%
\end{thebibliography}%
\begin{widetext}

\section{Supporting Information for
"Nonlinear circular dichroism in Mie-resonant nanoparticle dimers"}

Here, we provide more rigorous theoretical considerations to explain the circular dichroism, which appears in second harmonic signal of AlGaAs dimer with arbitrarily oriented crystalline lattice. All functions are written in a cylindrical coordinate system. Note, that when we speak about symmetry behavior of vector functions and tensors, we should take into account the behavior of basis vectors $\vec e_r, \vec e_z, \vec e_\varphi$. This makes considerations more sophisticated than in the scalar case. However, when considering vector functions as a whole (not each component separately), as it is usually done, for example, for vector spherical functions \cite{10.2307/43634833,ContributorstoWikimediaprojects2021-Vectorsphe,Zhang2008-Additiontheoremfor}, one can see, that everything is very similar to a scalar case. 
Namely, the  $\vec e_\varphi$ component is always written separately from others in our vectorial considerations, while it behaves diffetently under $\varphi\rightarrow-\varphi$ symmetry transformation, but vector fields as a whole have exactly the same symmetry behavior as a scalar functions given in the main text.

\subsection{Single dielectric resonator}


For the further convenience, let us use the cylindrical coordinate system for a single particle case, and assume that all fields are decomposed into the following series of cylindrical waves \cite{Venkatesh_Schurig_2016}:

\begin{equation}
 \vec E (r, \phi, z)=\sum_{m=-\infty}^{m=\infty} \vec E_m (r,z) e^{i m \vphi}.
\end{equation}
Normally incident circularly polarized plane wave with  contains only one term with $m=1$ (LCP) or $m=-1$ (RCP): 
\begin{equation}
\vec E^{wave}=E_x(\vec {e}_x\pm i \vec {e}_y)=E_{x}( \vec {e}_{r}\pm i  \vec {e}_{\varphi}) e^{\pm i \vphi}.
\end{equation} 
 Thus, only modes with $m=\pm 1$ are excited in a single cylinder, and
the field inside the cylinder:
\begin{gather} 
\vec E^{inc}=(E_{r}(r, z) \vec {e}_{r}+E_{z}(r, z) \vec {e}_{z}\pm i  E_{\varphi}(r, z)\vec {e}_{\varphi}) e^{\pm i \vphi}.
\end{gather}
This can be derived from the exact form of vector spherical harmonics \cite{Bohren}.
The second-harmonic polarization is written as following:
\begin{equation}
 \vec  P^{2\omega} (r, \phi, z)=\varepsilon_0 \hat\chi^{(2)} \vec  E^{inc} (r, \phi, z)\vec  E^{inc} (r, \phi, z)\label{pol}
\end{equation}
Then, let us consider the susceptibility tensor. 

For the AlGaAs ($[100]||x, \ [101]||y, \ [001]||z$), the nonlinearity tensor has only one independent component, ${\chi}_{xyz}={\chi}_{yxz}={\chi}_{xzy}={\chi}_{yzx}={\chi}_{zxy}={\chi}_{zyx}=\chi$.
{We can write it in the form 

\begin{equation}
 \hat{\chi}_{ijk}={\chi}_{ijk} \vec  e_i \vec  e_j \vec  e_k
\end{equation}
where Einstein summation convention is used.
Then, using the relations for the cylindrical coordinate system 

\begin{equation}\begin{cases}
\vec {e}_{x}=\cos\varphi\vec {e}_{r} - \sin\varphi\vec {e}_{\varphi}, \\
\vec {e}_{y}=\sin\varphi\vec {e}_{r} + \cos\varphi\vec {e}_{\varphi}, \\
\vec {e}_{z}=\vec {e}_{z},
\end{cases}\end{equation}
we write the $\hat{\chi}^{(2)}$ in cylindrical coordinates, for example, the following sum is 
\begin{equation}
\begin{aligned}
\vec {e}_{z}\vec {e}_{y}\vec {e}_{x}+\vec {e}_{z}\vec {e}_{x}\vec {e}_{y}= \\ =
{e^{2i\varphi}}\left(\frac1{2i}\vec {e}_{z}\vec {e}_{r}\vec {e}_{r}-\frac1{2i}\vec {e}_{z}\vec {e}_{\varphi}\vec {e}_{\varphi}+
\frac1{2}\vec {e}_{z}\vec {e}_{\varphi}\vec {e}_{r}+\frac1{2}\vec {e}_{z}\vec {e}_{r}\vec {e}_{\varphi}\right)+ \\+
{e^{-2i\varphi}}\left(-\frac1{2i}\vec {e}_{z}\vec {e}_{r}\vec {e}_{r}+\frac1{2i}\vec {e}_{z}\vec {e}_{\varphi}\vec {e}_{\varphi}+
\frac1{2}\vec {e}_{z}\vec {e}_{\varphi}\vec {e}_{r}+\frac1{2}\vec {e}_{z}\vec {e}_{r}\vec {e}_{\varphi}\right),
\end{aligned}
\end{equation}
the whole tensor for AlGaAs has four additional terms, $\vec {e}_{y}\vec {e}_{x}\vec {e}_{z}+\vec {e}_{x}\vec {e}_{y}\vec {e}_{z}+\vec {e}_{y}\vec {e}_{z}\vec {e}_{x}+\vec {e}_{x}\vec {e}_{z}\vec {e}_{y}$ which look similarly. 
Note, that nonlinear susceptibility tensor depends on $\varphi$, but only has two components  - $m=2$ and $m=-2$. We are only interested in a tensor's symmetry behavior. \cite{Dresselhaus,Ivchenko2012-SuperlatticesandOth,PhysRevB.99.075425} }

First, we consider  $[100]||x$- oriented lattice ($\beta=0$).

Substituting incident field inside the nanoparticle with  $m=\pm 1$ and polarizability tensor into Eq. \eqref{pol}, we obtain, that polarization can be written as

\begin{equation}
\begin{aligned}
 \vec P^{2\omega} (r, \phi, z)=\varepsilon_0 \hat\chi^{(2)} \vec E_1^{inc} \vec E_1^{inc} \propto \vec P_{e0}^{2\omega}  (r, z)\pm i\vec P_{o0}^{2\omega} (r, z)+\left(\vec P_{e4}^{2\omega}  (r, z)\pm i \vec P_{o4}^{2\omega}(r, z)\right) e^{ \pm 4 i \varphi}\label{pol2}
 \end{aligned}
\end{equation}
where the tensor gives its $m=2$ and $m=-2$  contribution and $\vec  P_{^e_o0}^{2\omega}  (r, z)$ and $\vec  P_{^e_o4}^{2\omega}  (r, z)$ can be assumed real, and $\pm$ in $e^{ \pm 4 i \varphi}$ refers to two different polarizations. Indexes $e$ and $o$ refer to $\varphi \rightarrow -\varphi$ parity, i.e. $\vec  P_{o0} \sim \vec  e_{\varphi}$,  $\vec  P_{e0} \sim \vec  e_{z}, \vec  e_{r}$.
%

If we rotate the crystalline lattice around the z-axis by the angle $\beta$, the induced polarization will be also rotated by the same angle and also will get an additional phase, which we are not interested in.

Thus, for the rotated lattice  $\varphi \rightarrow \varphi-\beta$:
\begin{equation}
 \vec P^{2\omega} (r, \phi, z)\propto  \vec P_{^e_o0}^{2\omega}  (r, z)+\vec P_{^e_o4}^{2\omega}  (r, z)e^{\pm 4 i (\varphi-\beta)}, \label{polar}
\end{equation} where $\vec P_{^e_o}=\vec P_{e}\pm i\vec P_{o}$.

To obtain the important conclusions, we don't need to know how exactly the eigenmodes and polarization look like, we only need to know, how they depend on $\varphi$. 
\paragraph{Cylinder modes.}
Each mode corresponds to particular $|m|$. Modes with different $|m|$ have different eigenfrequency, thus the resonant-state expansion\cite{PhysRevA.90.013834} will look as following:
\begin{equation}
\begin{aligned}
\hat {{\bf G}}({\bf r,r'},k)=\sum_{m,n} \frac{\mathbf{E}_{emn}(\mathbf{r}) \otimes \mathbf{E}_{emn}\left(\mathbf{r}^{\prime}\right)+\mathbf{E}_{omn}(\mathbf{r}) \otimes \mathbf{E}_{omn}\left(\mathbf{r}^{\prime}\right)}{2 k\left(k-k_{mn}\right)},
\end{aligned}
\end{equation}
where $n$ is number of mode. Importantly, even and odd modes are degenerated. Eigenvectors $k_{mn}$ are complex.
Under circularly polarized plane wave only modes with $m=0$ and $m=4$ are excited in SH, because all the other overlap integrals are zero. Modes with $m=0$ and $m=4$ will be excited with different phases. After all, the SH field will be written as following:
\begin{equation}
\mathbf{E}^{2\omega}({\mathbf{r}})=\sum_{n,\nu} \left( g_n(\omega) \vec  E_{^e_o0n}^{2\omega}+q_\nu(\omega) \vec  E_{^e_o4\nu}^{2\omega}e^{\pm 4i(\varphi-\beta)}\right)
\end{equation}
where $n$ and $\nu$ is number of a mode, and  $g_n(\omega)$ and $q_{\nu}(\omega)$ are complex coefficients. 


\subsection{Nanoparticle dimer}
Modes of a dimer can be considered with help of multipole decomposition\cite{Gladyshev_Frizyuk_Bogdanov_2020} in two different ways:

1) Every particle has its own multipole decomposition, as shown in Fig. 3(a) of the main text. This approach is useful for mode hybridization consideration. However, in Fig. 3(a), only the main contribution is shown, actually, there will be an infinite row of multipoles. 

2) Consideration of the multipolar content of the far-field. In this approach, only symmetry of the dimer matters, so the multipolar content will be the same as for a single elliptical along the $x$-axis particle, for example. This approach helps to understand, that all modes, which are transformed by the same irreducible representation, have the same multipole content, and can add up in the overall intensity.


How the dimerization affect our fields and intensity?

Now, we will use the first approach to describe the modes and fields in each particle. 
    Thus, after dimerization, the modes of sinle particle are perturbed as following:
\begin{equation}
\begin{aligned}
\vec  E_{0n}^{2\omega}\rightarrow \vec  E_{e0n}^{2\omega}+\vec  E_{e0n4}^{2\omega}\cos(4  \varphi)+\vec  E_{o0n4}^{2\omega}\sin(4  \varphi)+\dots=E_{0r}\vec e_r+E_{0z}\vec e_z+\dots \label{mode1}
\end{aligned}
\end{equation}
\begin{equation}
\begin{aligned}
\vec  E_{4n}^{2\omega}\rightarrow \vec  E_{e4n4}^{2\omega}\cos(4  \varphi)+\vec E_{o4n4}^{2\omega}\sin(4  \varphi)+\vec  E_{e4n0}^{2\omega}+\dots
=(E_{4r}\vec e_r+E_{4z}\vec e_z)\cos(4\varphi)+E_{4\varphi}\vec e_\varphi\sin(4\varphi)+\dots\label{mode2}
\end{aligned}
\end{equation}
for even modes ($A_1$), and
\begin{equation}
\begin{aligned}
\vec  E_{0n}^{2\omega}\rightarrow \vec  E_{o0n0}^{2\omega}+\vec  E_{e0n4}^{2\omega}\sin(4  \varphi)+\vec  E_{o0n4}^{2\omega}\cos(4  \varphi)+\dots=E_{0\varphi}\vec e_\varphi+\dots \label{mode3}
\end{aligned}
\end{equation}
\begin{equation}
\begin{aligned}
\vec  E_{4n}^{2\omega}\rightarrow \vec  E_{e4n4}^{2\omega}\sin(4  \varphi)+\vec  E_{o4n4}^{2\omega}\cos(4  \varphi)+\vec  E_{o4n0}^{2\omega}+\dots=(E_{4r}\vec e_r+E_{4z}\vec e_z)\sin(4\varphi)+E_{4\varphi}\vec e_\varphi\cos(4\varphi)+\dots \label{mode4}
\end{aligned}
\end{equation}
for odd modes ($A_2$). 
Infinite sums containing all possible multipoles, even (Eq. \eqref{mode1},  \eqref{mode2})  or odd (Eq. \eqref{mode3},  \eqref{mode4}) under $\varphi\rightarrow-\varphi$ transformation, but below, we only consider dominating terms.

Let us write the overlap integral  between dimer modes and polarization. 
Let us consider two $A_1$ modes \eqref{mode1},  and  \eqref{mode2}. 
We write the integral only over one cylinder (coordinate system located in center of the cylinder), because over the second will be the same:


{\begin{equation}
\begin{aligned}
D_0=\int\limits_V (E_{0r}\vec {e}_{r}+E_{0z} \vec {e}_{z}) \vec P_{^e_o0}^{2\omega}(r, z) \propto 2\pi,
\end{aligned}
\end{equation}}
{\begin{equation}
\begin{aligned}
D_4=\int\limits_V dV\left((E_{4r}\vec {e}_{r}+E_{4z} \vec {e}_{z})\cos(4\varphi)+ E_{4\varphi}\vec {e}_{\varphi}\sin(4\varphi)\right)\cdot \left(\vec P_{^e_o4}^{2\omega}(r, z)e^{\pm 4 i (\varphi-\beta)}\right)\propto e^{\mp 4i\beta}
\end{aligned}
\end{equation}}

which coincides with formulae (5) and (6) in the main text.  Second approach (Fig.3 (b), (c)) helps us to understand, why 
the sum of $m=0$ and $m=4$ modes lead to CD appearance. Both of them belong to the $A_1$ symmetry, hence, they have the same multipolar content in the far-field. Further considerations do not require vector functions, and provided in the main text. 
\end{widetext}






\end{document}